\documentclass[useAMS,usenatbib]{mn2e}

\usepackage{amsmath}
\usepackage{amssymb}
\usepackage{graphicx,epsfig}
\usepackage{times}

\def\gsim{\;\lower4pt\hbox{${\buildrel\displaystyle >\over\sim}$}\;}
\def\lsim{\;\lower4pt\hbox{${\buildrel\displaystyle <\over\sim}$}\;}
\def\grls{\;\lower4pt\hbox{${\buildrel\displaystyle >\over <}$}\;}
\def\beq{\begin{equation}}
\def\eeq{\end{equation}}

\title[$M_{\rm bh}$-$\sigma_*$ relation: bulges vs. pseudobulges]{The black hole mass -- stellar velocity dispersion correlation:\\ bulges versus pseudobulges}
\author[J. Hu]{Jian Hu\thanks{Email: jhu@mpa-garching.mpg.de}\\
Max-Planck-Institut f\"ur Astrophysik, Karl-Schwarzschild-Stra\ss e 1, D-85741 Garching bei M\"unchen, Germany}

\begin{document}

\date{Accepted 2008 ... Received 2008 ...; in original form 2008 ...}

\pagerange{\pageref{firstpage}--\pageref{lastpage}} \pubyear{2008}

\maketitle

\label{firstpage}

\begin{abstract}

We investigate the correlation between the supermassive black holes (SMBHs) mass ($M_{\rm bh}$) and the stellar velocity dispersion ($\sigma_*$) in two types of host galaxies: the early-type bulges (disk galaxies with classical bulges or elliptical galaxies) and pseudobulges. In the form $\log (M_{\rm bh}/{\rm M_\odot})=\alpha+\beta\log(\sigma_*/200\ {\rm km\ s^{-1}})$, the best-fit results for the 39 early-type bulges are the slope $\beta=4.06\pm0.28$ and the normalization $\alpha=8.28\pm0.05$; the best-fit results for the 9 pseudobulges are $\beta=4.5\pm1.3$, $\alpha=7.50\pm0.18$. Both relations have intrinsic scatter in $\log M_{\rm bh}$ of $\lesssim$0.27 dex. The $M_{\rm bh}$-$\sigma_*$ relation for pseudobulges is different from the relation in the early-type bulges over the 3$\sigma$ significance level. The contrasting relations indicate the formation and growth histories of SMBHs depend on their host type. The discrepancy between the slope of the $M_{\rm bh}$-$\sigma_*$ relations using different definition of velocity dispersion vanishes in our sample, a uniform slope will constrain the coevolution theories of the SMBHs and their host galaxies more effectively. We also find the slope for the ``core'' elliptical galaxies at the high mass range of the relation appears steeper ($\beta\simeq$ 5-6), which may be the imprint of their origin of dissipationless mergers. 

\end{abstract}

\begin{keywords}
black hole physics -- galaxies: bulges -- galaxies: formation -- galaxies: fundamental parameters -- galaxies: nuclei. 
\end{keywords}

\section{Introduction}

Supermassive black holes (SMBHs) are ubiquitous in the center of elliptical galaxies and bulges of disk galaxies.
The masses ($M_{\rm bh}$) of the nearby mostly quiescent SMBHs are found to be correlated with the physical properties of their host bulges, such as the luminosity $L$ (Kormendy \& Richstone 1995; Marconi \& Hunt 2003), the mass $M_{\rm b}$ (Magorrian et al. 1998; Marconi \& Hunt 2003; H\"aring \& Rix 2004), the stellar velocity dispersion $\sigma_*$ (Ferrarese \& Merritt 2000; Gebhardt et al. 2000a; Tremaine et al. 2002; Ferrarese \& Ford 2005), the galaxy light concentration $C_{\rm Re}$ (Graham et al. 2001), the S\'ersic (1968) index $n$ of surface brightness profile (Graham \& Driver 2007), and the inner core radius $r_\gamma$ (Lauer et al. 2007a). 
Among these, the  $M_{\rm bh}$-$\sigma_*$ relation of the form 
\begin{equation}
\log(M_{\rm bh}/{\rm M_\odot})=\alpha+\beta\log(\sigma/200\ {\rm km\ s^{-1}})
\end{equation}
is one of the most tight with a intrinsic scatter of $\lesssim$0.3 dex (Novak et al. 2006). The validity of this relation have been explored for intermediate mass black holes (IMBHs) in dwarf galaxies (Barth et al. 2005), the local active galaxies (McLure \& Dunlop 2002; Green \& Ho 2006), and Seyfert galaxies at $z=0.36$ (Woo et al. 2006; Treu et al. 2007). 
In fact, this relation was predictd by Silk \& Rees (1998), Fabian (1999) and Blandford (1999). 
As a crucial test of theoretical models of formation and coevolution of the SMBHs and their host galaxies, many efforts have been made to explain the relation since its recognition (e.g. King 2003 and reference therein). The normalization, intrinsic scatter, and redshift-dependent evolution of the relation are important for study of SMBHs demography (e.g. Yu \& Tremaine 2002; Shankar et al. 2004). 

The early published estimation of the slope $\beta$ are diverse from the highest value $4.8\pm0.5$ (Ferrarese \& Merritt 2000, hereafter FM00) to the lowest value $3.75\pm0.3$ (Gebhardt et al. 2000a, hereafter G00). Tremaine et al. (2002, hereafter T02) explored the reasons for the discrepancy between the results of the two groups, and show the systematic differences in the velocity dispersion measurements contribute the main diversity in the slope. The G00 and T02 define ``effective stellar velocity dispersion'' $\sigma_{\rm e}$ as the luminosity-weighted rms dispersion within a slit aperture of length $2R_{\rm e}$, while FM00 use ``central stellar velocity dispersion'' $\sigma_{\rm c}$ normalized to an aperture of radius $R_{\rm e}/8$, where $R_{\rm e}$ is the effective radius of the galaxy bulges. 
T02 presented a best-fit result of the  $M_{\rm bh}$-$\sigma_{\rm e}$ relation with $\alpha=8.13\pm0.06$ and $\beta=4.02\pm0.32$, which is extensively used in the subsequent studies.
The most current version of the $M_{\rm bh}$-$\sigma_{\rm c}$ relation are $\alpha=8.22\pm0.06$ and $\beta=4.86\pm0.43$ (Ferrarese \& Ford 2005, hereafter FF05). 
Although both of the relations are tight, their slopes still differ by 1.6 standard deviations. 

Besides the uncertainty in slope and normalization, the dependence of the  $M_{\rm bh}$-$\sigma_*$ relation on the type of SMBHs host is also worth consideration. 
A variety of observational and theoretical results show that there are two kinds of ``bulges'' in disk galaxies\footnote{Athanassoula (2005) distinguished three types of bulges: classical bulges, boxy/peanut bugles and discy bulges. The boxy/peanut bugles are in fact a part of the bar seen edge-on; the discy bulges result from the inflow of gas and star formation in the central region of disk galaxies. Terminologically, the ``pseudobulges" in this paper include boxy/peanut bugles and discy bulges, as both of them are due to secular evolution and unrelated with the major mergers.} (e.g. Kormendy \& Kennicutt 2004, hereafter KK04). The classical bulges dwelling in the center of disk galaxies are like little elliptical galaxies, which are thought to be the products of violent relaxation during major mergers. On the other hand, pseudobulges in disk galaxies are physically unrelated to ellipticals, their structure and kinematics resemble that of disks. They are believed to have formed via distinct scenario than that of classical bulges. There is no guarantee that SMBHs in pseudobulges should follow the same  $M_{\rm bh}$-$\sigma_*$ relation in ellipticals. In contrast, several evidences suggest they probably obey a different relation. (i) The rotationally flattened pseudobulges have smaller velocity dispersion than predicted by the Faber-Jackson $\sigma_*$-$L$ correlation (e.g. KK04). If the pseudobulge follow the same $M_{\rm bh}$-$L$ relation, they will naturally deviate from the same  $M_{\rm bh}$-$\sigma_*$ relation. (ii) The pseudobulges are dominantly shaped by the secular evolution, and probably never undergo a major merger, which is postulated as the key mechanism to explain the $M_{\rm bh}$-$\sigma_*$ relation in the ellipticals. The gas inflow fueling SMBHs induced by the secular evolution is not as effective as that in major merger events. 
(iii) The slope of the  $M_{\rm bh}$-$\sigma_*$ relation in the local active galactic nuclei (AGNs) flattens somewhat at the low mass end (Greene \& Ho 2006). Because most of the local low mass AGNs have late-type hosts (Kauffmann et al. 2003), we speculate some (or even most) of these low mass SMBHs may live in the pseudobulges, and disagree with the relation for high mass AGNs with (mostly) early-type hosts. 

Kormendy \& Gebhardt (2001) checked the $M_{\rm bh}$-$\sigma_*$ relation for five pseudobulges, and found no obvious inconsistency with the relation for the ellipticals. However, one of these galaxies (NGC 4258) was in fact a classical bulge, incorrectly identified as a pseudobulge (J. Kormendy, private communication), and is consistent with the $M_{\rm bh}$-$\sigma_*$ relation for the elliptical galaxies, while the $M_{\rm bh}$ of the other four pseudobulges appears smaller than the prediction by the relation (cf. Fig. 5 of Kormendy \& Gebhardt 2001). As pointed out by the authors, it is important to check their result with a larger sample. 
Therefore, in this paper we investigate the $M_{\rm bh}$-$\sigma_*$ relation with the up-to-date sample of early-type bulges and pseudobuges. 

The paper is organized as follows. In section 2, we describe the sample used for analysis. In section 3, we determine and compare the  $M_{\rm bh}$-$\sigma_*$ relation in early-type bulges and pseudobulges. In section 4, we summarize and discuss our results.
Throughout this paper, we use the the base 10 logarithms, and cosmological constants $\Omega_{\rm M}=0.3$, $\Omega_\Lambda=0.7$, $h=0.7$.

\begin{table*}
 \centering
\begin{minipage}{170mm}
  \caption{The sample of black holes in classical bulges and elliptical galaxies.}
  \begin{tabular}{l l c c c l c c l c c l}
  \hline
Galaxy & Type & Activity & $D$ & $M_{\rm bh}$ & ref & $\sigma_{\rm e}$ & $\sigma_{\rm c}$  & ref & $n$ & $R_{\rm e}$ & ref 
\\
 & & &(Mpc) & (10$^8$ M$_\odot$) &  & ($\rm km\ s^{-1}$) & ($\rm km\ s^{-1}$) & & & (kpc) &
\\
(1) & (2) &(3) &(4) &(5) &(6) &(7) &(8) &(9) &(10) &(11) &(12)
\\
\hline
N221
& E2 & n & 0.81 & $0.025\pm0.005$ &  s-45
& $75\pm4$ & $72\pm4^{\rm a}$ & 43 
& 3.7 
& 0.24  &  37, 32 
\\
N224
& Sb & L & 0.76& $1.4^{+0.9}_{-0.3}$ & s-4 
& $160\pm8$ & $160\pm8$ & 43  
& 2.2 
& 1.0 & 29
\\
N821 & E4 &n&	24.1 & $0.85\pm0.35$ & s-38
& $189\pm9$ & $203\pm10^{\rm a}$
&  8
& 4.0 
& 5.23 & 16  
\\
N1023 &	SB0 & n & 11.4 & $0.44\pm0.05$ & s-11
& $205\pm10$ & $205\pm10^{\rm a}$ 
& 11
& 2.0 
& 2.7 &  23, 8
\\
N1399$^{\rm c}$ &	E1 & n & 21.1 & $12^{+5}_{-6}$ & s-28
& $317\pm16$ & $329\pm16^{\rm a}$ 
& 28
& 4.9 
& 4.1 &  39
\\
N2974 & E4 & Sy2& 21.5 & $1.7\pm0.2$& s-9
& $233\pm12$ & $236\pm12^{\rm a}$ 
& 8
& 4.1  
& 2.5 & 37, 8
\\
N3031
& Sab&L & 3.9 & $0.8^{+0.2}_{-0.1}$ & g-14
& $173\pm17$  
& $173\pm17$
& 47
& 2.0 
& 0.21 & 30 
\\
N3115 & S0& n & 9.7 & $10^{+10}_{-6}$ & s-15
& $230\pm12$
& $256\pm13^{\rm a}$ 
 & 43 
& 4.4  
& 4.7 &  37, 32 
\\
N3245 & S0 & L & 20.9 & $2.1\pm0.5$ & g-2
& $205\pm10$	& $222\pm11^{\rm a}$ 
& 43
& 4.0 
& 1.08 & 16 
\\
N3377 &	E5 & n &	11.2	& $0.7\pm0.1$ & s-10
& $138\pm7$ & $145\pm7^{\rm a}$ 
&  8 
& 3.0 
& 2.38 & 16
\\
N3379$^{\rm c}$
& E1& L & 10.3 & $1.2\pm0.6$ & s,g-21
& $201\pm10$	&$217\pm11^{\rm a}$ 
&  8
& 4.3 
& 2.2 & 23, 8
\\
N3414 & S0& n & 25.2 & $2.5\pm0.3$& s-9
& $205\pm10$ & $242\pm12^{\rm a}$ 
&  8
& 3.8  
& 4.0 & 37, 8
\\
N3608$^{\rm c}$ & E2 & L & 22.9 & $1.9^{+1.0}_{-0.6}$ & s-22
& $178\pm9$	& $196\pm10^{\rm a}$
&  8
& 4.9  
&4.6 & 37, 8 
\\
N3998 & S0 & L & 18.3 & $2.9^{+0.3}_{-0.5}$ & g-12
& $268\pm14$ & $268\pm14$ & 46
 & 4.1 
& 0.70 & 12
\\
N4151 & Sa & Sy1& 13.9 & $0.3^{+0.1}_{-0.2}$ & s-36,g-27
& $97\pm10$ & $97\pm10$ &  34
& 3.0 &  0.32 & 20
\\
N4258 & SBbc & L &7.2 & $0.39\pm0.01$ &	m-26
& $148\pm7$	& $120\pm6$ & 3   
& 3.0 
& 3.1 & 41
\\
N4261$^{\rm c}$ & E2& L & 31.6 &	$5.2^{+1.0}_{-1.1}$	& g-18 
& $309\pm15^{\rm a}$ & $309\pm15^{\rm a}$
&  
& 3.8 
& 6.5  & 37, 32 
\\
N4291$^{\rm c}$ & E2 & n & 26.2 &	$3.1^{+0.8}_{-2.3}$	& s-22 
& $242\pm12$&  $249\pm12^{\rm a}$ 
& 22
& 4.6 
& 2.04 & 16
\\
N4459 & S0 & L & 16.1&  $0.70\pm0.13$ & g-40
& $168\pm8$ & $178\pm9^{\rm a}$ 
& 8  
& 3.5  
& 3.0 &  37, 8
\\
N4473$^{\rm c}$	& E5 & n &15.3 & $1.1^{+0.4}_{-0.8}$ & s-22 
&$192\pm10$ & $188\pm9^{\rm a}$ 
& 8  
& 2.7 
& 2.1 &  23, 8
\\
N4486$^{\rm c}$
& E0 & L & 17.2 & $30\pm10$ & g-24
& $298\pm15$	& $329\pm16^{\rm a}$ 
& 8 
& 6.9 
& 8.8 & 23, 8
\\
N4486A & E2 & n & 18.3 & $0.13^{+0.04}_{-0.07}$ & s-35	
&$110\pm10^{\rm b}$		
& $110\pm10^{\rm b}$ 
&  
& 2.2 
& 0.54 &  35, 17 
\\
N4552$^{\rm c}$
& E & L & 15.9 & $5.0\pm0.5$& s-9
& $252\pm13$
& $263\pm13^{\rm a}$ 
 &  8 
 & 4.8 
& 2.4& 37, 8  
\\
N4564 & E6 &	n&15.9 &	$0.59^{+0.03}_{-0.08}$ & s-22 
& $162\pm8$ & $167\pm8^{\rm a}$ &  43 
& 3.2 
& 1.54 &  23, 22
\\
N4596 & SB0 &  L & 16.7 & $0.78^{+0.42}_{-0.33}$ & g-40
& $152\pm8$ & $149\pm8^{\rm a}$ 
&  40 
& 1.4 
&  0.23 & 31
\\
N4621
& E5& n &18.3 & $4.0\pm0.4$& s-9 
& $211\pm11$
& $231\pm12^{\rm a}$ 
 &  8 
 & 5.2 
&4.1& 37, 8
\\
N4649$^{\rm c}$ & E1	& n& 17.3	& $20^{+4}_{-6}$ & s-22
& $330\pm17^{\rm b}$	
& $337\pm17^{\rm a}$
&   %
& 6.0 
& 6.13 & 23, 22 
\\
N4697 & E4 &	n &11.7&	$1.7^{+0.2}_{-0.3}$ & s-22 
& $177\pm9$	
& $178\pm9^{\rm a}$ 
&  22
& 4.0 
&4.25&  23, 22
\\
N4742 & E4 & n  & 15.5 & $0.14^{+0.04}_{-0.05}$ & s-43 
& $90\pm5$ & $109\pm5^{\rm a}$ & 44 
& 5.9 
&  0.60 &  37 
\\
N5077$^{\rm c}$ & E3 & L & 40.2 & $7.2^{+4.5}_{-3.0}$ & g-13 
& $261\pm13^{\rm a}$ & $261\pm13^{\rm a}$ &
& 4.7 
& 3.8 & 13
\\
N5128 & S0 &Sy2& 3.5 &	$0.45^{+0.06}_{-0.04}$ &	g-25	
& $138\pm10$ & $128\pm8^{\rm a}$ 
& 43  
& 4.0 & 1.4 & 42 
\\
N5252 & S0 & Sy2 
& 96.8 & $10^{+15}_{-5}$ & g-5
& $190\pm27$&$190\pm27$ & 33
& 6.0 
&  9.7 & 6, 32 
\\
N5813$^{\rm c}$ & E1& L& 32.2 & $7.0\pm0.7$& s-9
& $230\pm12$ 
& $237\pm12^{\rm a}$
&  8 
& 4.9 
&8.1 & 37, 8
\\
N5845 & E3& n & 25.9	& $2.4^{+0.4}_{-1.4}$ & s-22
& $239\pm12$
& $258\pm13^{\rm a}$
 & 8  
 & 3.2
& 0.58 & 23, 8 
\\
N5846 & E0 & n & 24.9 & $11\pm1$& s-9
& $238\pm12$ & $236\pm12^{\rm a}$ 
&   8 
& 3.9 
& 9.8 & 37, 8  
\\
N6251 & E2& Sy2 & 107 & $6.1^{+2.0}_{-2.1}$ & g-19 
& $290\pm15$ & $305\pm21^{\rm a}$ 
&  43 
&4.0 
& 11 & 32 
\\
N7052$^{\rm c}$	& E4 & L	& 71.4 &	$4.0^{+2.8}_{-1.6}$	& g-44
& $266\pm13$	
& $265\pm13^{\rm a}$ 
&  44
& 4.6 
& 12 &  23, 32
\\
N7457 &	S0 & n & 13.2	& $0.035^{+0.011}_{-0.014}$ & s-22 
& $78\pm4$ & $80\pm5^{\rm a}$ & 8 
& 2.0 
& 4.2  &  23, 8
\\
IC 1459$^{\rm c}$ & E3&L &	29.2 & $25^{+5}_{-4}$ &s-7
& $340\pm17$	& $304\pm15^{\rm a}$ 
& 7 
& 4.9 
&  8.2 &  37, 32
\\
\hline
\end{tabular}
\\
Notes: (1) name of the galaxy. (2) Hubble type of the galaxy. (3) the activity of the galactic nuclei (n = inactive; Sy1 = Seyfert 1; Sy2 = Seyfert 2; L = LINER). (4) distance to the galaxy. (5) mass of the central black hole. (6) method (s = stellar kinematics; g = gas kinematics; m = H$_2$O masers; p = stellar proper motions) and reference for the black hole mass measurement. (7-9) effective and central stellar velocity dispersion of the bulge, and their references. (10-12) S\'ersic index and effective radius of the bulge surface brightness profile, and their references.

$^{\rm a}$ The value is taken from HyperLeda database, corrected with the aperture of $R_{\rm e}$, the error is taken as $5\%$ for the E/SO galaxies and $10\%$ for the Sa-Sd galaxies.

$^{\rm b}$ The value is estimated by the author, see details in section 2.2.

$^{\rm c}$ The ``core'' elliptical galaxies (Lauer et al. 2007a).

References: 
(1) Balcells et al. 2007; 
(2) Barth et al. 2001; 
(3) Batcheldor et al. 2005; 
(4) Bender et al. 2005; 
(5) Capetti et al. 2005; 
(6) Capetti \& Balmaverde 2007; 
(7) Cappellari et al. 2002; 
(8) Cappellari et al. 2006; Cappellari et al. 2007a;
(9) Cappellari et al. 2007b; 
(10) Copin et al. 2004; 
(11) Bower et al. 2001;
(12) De Francesco et al. 2006; 
(13) De Francesco et al. 2008; 
(14) Devereux et al. 2003; 
(15) Kormendy et al. 1996; 
(16) Erwin et al. 2002; 
(17) Faber et al. 1989; 
(18) Ferrarese et al. 1996; 
(19) Ferrarese \& Ford 1999; 
(20) Gadotti 2008;
(21) Gebhardt et al. 2000b; Shapiro et al. 2006; 
(22) Gebhardt et al. 2003;	
(23) Graham \& Driver 2007; 
(24) Harms et al. 1994; Macchetto et al. 1997; 
(25) Neumayer et al. 2007; 
(26) Herrnstein et al. 1999;
(27) Hicks \& Malkan 2008; 
(28) Houghton et al. 2006; 
(29) Kormendy \& Bender 1999; 
(30) Laurikainen et al. 2004; 
(31) Laurikainen et al. 2005; 
(31) Bower et al. 2001;
(32) Marconi \& Hunt 2003;
(33) Nelson \& Whittle 1995; 
(34) Nelson et al. 2004; 
(35) Nowak et al. 2007; 
(36) Onken et al. 2007; 
(37) Prugniel \& H\'eraudeau 1998; 
(38) Richstone et al. 2004; 
(39) Saglia et al. 2000; 
(40) Sarzi et al. 2001; 
(41) Scarlata et al. 2004; 
(42) Silge et al. 2005; 
(43) Tremaine et al. 2002; 
(44) van der Marel \& van den Bosch 1998; 
(45) Verolme et al. 2002; 
(46) Wegner et al. 2003;
(47) Vega Beltr\'an et al. 2001.
\end{minipage}
\end{table*}
\begin{table*}
 \centering
\begin{minipage}{170mm}
  \caption{The sample of black holes in pseudobulges.}
  \begin{tabular}{l l c c c l c c  l c c l}
    \hline
Galaxy & Type & Activity & $D$ & $M_{\rm bh}$ & ref & $\sigma_{\rm e}$ & $\sigma_{\rm c}$  & ref & 
$n$ & $R_{\rm e}$ & ref 
\\
 & & &(Mpc) & (10$^8$ M$_\odot$) &  & ($\rm km\ s^{-1}$) & ($\rm km\ s^{-1}$) & & & (kpc) &
\\
(1) & (2) &(3) &(4) &(5) &(6) &(7) &(8) &(9) &(10) &(11) &(12) 
\\
\hline
N1068 & SBb &Sy2 &15.3 & $0.083\pm0.003$ & m-12 
& $165\pm17$ & $165\pm17$  & 13 
& 1.7  
& ... & 3
\\
N2787 &	SB0 & L  &7.5	& $0.41^{+0.04}_{-0.05}$ & g-14
& $218\pm11$ & $218\pm11$ 
& 1 
& 1.0 
&  0.32 & 4 
\\
N3079 & SBcd & Sy2& 19.1 & $0.025_{-0.013}^{+0.025\rm b}$ & m-18
& $146\pm15$ & $146\pm15$ & 20 
&... & ...&
\\  
N3227 & SBa & Sy1 &17.5 & $0.20^{+0.14}_{-0.13}$ & s,g-2
& $167\pm17$ &  $167\pm17$ & 2
& 1.1 
&  0.31 &  5 
\\
N3384 &	SB0 & n & 11.6	& $0.16^{+0.01}_{-0.02}$	& s-7
& $143\pm7$ & $148\pm7^{\rm a}$ & 14
&  1.0 
& 1.2 &  4 
\\
N3393 & SBa & Sy2 & 51.8 & 0.31$\pm$0.02 & m-11 
& 184$\pm$18 & 184$\pm$18 & 15
&...&...&
\\
Circinus & Sb& Sy2 & 2.8 & 0.011$\pm$0.002 & m-10
&  $75\pm20$  & $75\pm20$ & 21
& ... &  ... & 
\\
IC 2560 & SBb & Sy2 & 41.4 & 0.029$\pm$0.006 & m-16
& 137$\pm$14 & 137$\pm$14 & 6 
& ... & 0.34 & 19   
\\
MW & SBbc & n & 0.008 & 0.037$\pm$0.002 & p-8 
& $103\pm20$ & $100\pm 20$  & 17
& 1.3 
& 0.7 & 9, 17 
\\
\hline
\end{tabular}
\\
Notes: Same as Table 1.

$^{\rm a}$ The value is taken from HyperLeda database, corrected with the aperture of $R_{\rm e}$ (if available), the error is taken as $5\%$ for the E/SO galaxies and $10\%$ for the Sa-Sd galaxies.

$^{\rm b}$ The mass uncertainty is not given in the literature, we use $\Delta\log M_{\rm bh}$=0.3 as a rough estimation.

References: 
(1) Bertola et al. 1995;
(2) Davies et al. 2006; Hicks \& Malkan 2007; 
(3) Drory \& Fisher 2007; 
(4) Erwin et al. 2002; 
(5) Gadotti 2008; 
(6) Garcia-Rissmann et al. 2005;
(7) Gebhardt et al. 2003; 
(8) Ghez et al. 2005;
(9) Graham \& Driver 2007; 
(10) Greenhill et al. 2003; 
(11) Kondratko et al. 2008; 
(12) Lodato \& Bertin 2003; Hur\'e 2002; 
(13) Nelson \& Whittle 1995; 
(14) Sarzi et al. 2001; 
(15) Terlevich et al. 1990; 
(16) Tilak et al. 2008; 
(17) Tremaine et al. 2002; 
(18) Yamauchi et al. 2004; Kondratko, Greenhill, \& Moran 2005;
(19) Fathi \& Peletier 2003; 
(20) Shaw et al. 1993;
(21) Maiolino et al. 1998.
\end{minipage}
\end{table*}

\section{The Sample}

We select the sample with spatially well-resolved $M_{\rm bh}$ measurement from literature, and classify them as early-type bulges or pseudobulges. 
The selection criteria is $2r_{\rm bh}/r_{\rm res}>$1, where $r_{\rm bh}=GM_{\rm bh}/\sigma_*^2$ is the radius of the SMBHs influence sphere, $r_{\rm res}$ is the instrumental spatial resolution. 

The properties of the bulges and SMBHs are listed in Table 1 and Table 2. Our sample (39 early-type bulges + 9 pseudobulges) is larger than T02 (31 galaxies) and FF05 (30 galaxies), with updated mass and/or velocity dispersion measurement. 
Among the 48 objects, 28 are weighted by stellar dynamics, 16 by gas kinematics (3 have both stellar and gas kinematics measurements), 6 by masers, 1 by stellar proper motions; 21 are inactive, 16 are LINERs, 2 are Seyfert 1, and 9 are Seyfert 2.  

Most distances of galaxies in the sample have been measured with the surface brightness fluctuation (SBF) method (Tonry et al. 2001). 
The distances to the galaxies in Virgo cluster (NGC 4473, NGC 4486, NGC 4486A, NGC 4552, NGC 4564, NGC 4649) are based on the new SBF measurement by Mei et al. (2007).
For those without an SBF distance we use recession velocities corrected for Virgocentric infall from HyperLeda database\footnote{http://leda.univ-lyon1.fr/}, except some individual galaxies explained in section 2.2. 
The black hole mass is linearly modified according to the distance that is different from the value used in the literature describing the mass measurement. The black hole mass errors are converted to 1$\sigma$ values if they are given at different confidence level (C.L.) in the literature.

\begin{figure}
\centerline{\includegraphics[width=8.5cm]{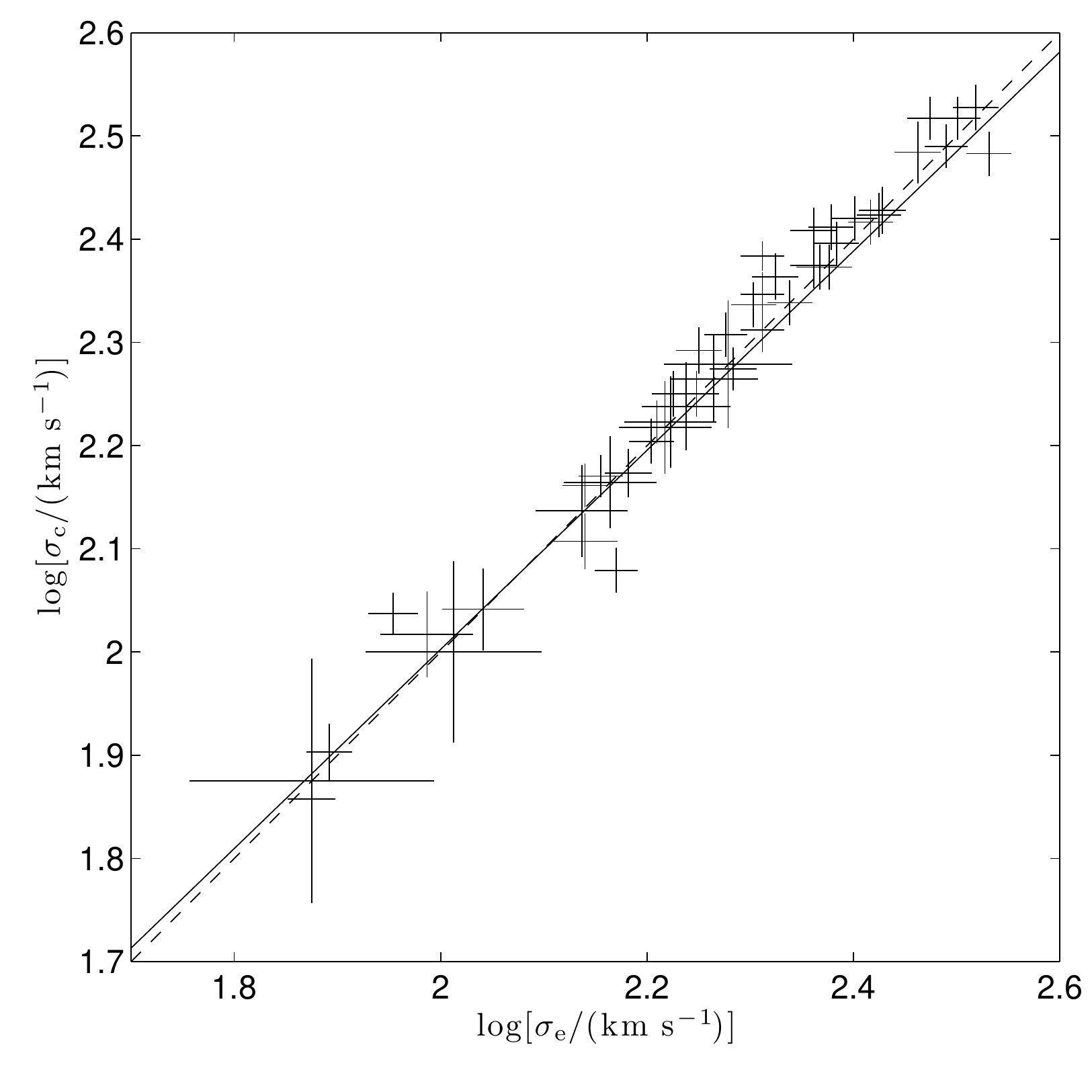}} 
\caption{Comparison of the $\sigma_{\rm e}$ and $\sigma_{\rm c}$ of the sample galaxies. The solid line denotes the model of T02 (eq. [2]). The dashed line denotes $\sigma_{\rm e}=\sigma_{\rm c}$.}
\end{figure}

Most of the velocity dispersions $\sigma_{\rm c}$ of E/S0 galaxies are calculated using the formula 
$\sigma_{\rm c}=\sigma_{\rm c0}(8R_{\rm ap}/R_{\rm e})^{0.04}$ (J\o rgensen, Franx, \& Kjaergaard 1995), 
where $\sigma_{\rm c0}$ are provided by the Hyperleda database, $R_{\rm ap}=0.85$ kpc, $R_{\rm e}$ is the effective radius of bulges surface brightness profile, listed in the column (11) of Table 1 and 2. 
For Sa-Sd galaxies, observed $\sigma_*$ profiles is not central peaked but nearly constant in the central $10''$ region, the variation of $\sigma_*$ measured with different apertures is small (Pizzella et al. 2004; Baes et al. 2003), so we take $\sigma_*$ value measured with an small aperture without further corrections.
We adopt $5\%$ error of $\sigma_{\rm e}$ and $\sigma_{\rm c}$ for E/S0 galaxies and $10\%$ error for Sa-Sd galaxies, except for some individual galaxies explained in section 2.2. 
For those without available $R_{\rm e}$ measurement, we simply use $R_{\rm e}=R_{\rm ap}=0.85$ kpc, which is supposed to be an acceptable approximation to the true value.
For those without $\sigma_{\rm e}$ in literature, we take the value of $\sigma_{\rm c}$, since they are very close on average. T02 have modeled the $\sigma_{\rm e}$-$\sigma_{\rm c}$ relation for 39 early-type galaxies, and found (cf. eq. [16] in T02):
\beq
\log\Big(\frac{\sigma_{\rm c}}{\sigma_{\rm e}}\Big)=(-0.008\pm0.004)-(0.035\pm0.017)\log\Big(\frac{\sigma_{\rm c}}{200\ {\rm km\ s^{-1}}}\Big).
\eeq
Comparison of the $\sigma_{\rm e}$ and $\sigma_{\rm c}$ of our sample galaxies is shown in Figure 1. 
Obviously, the difference between the $\sigma_{\rm e}$ and $\sigma_{\rm c}$ are much smaller than their measurement errors. 

\subsection{Identification of pseudobulges}

A comprehensive review on the pseudobulge properties and methods to recognize pseudobulges are tabulated in Kormendy \& Kennicutt (2004). 
S\'ersic index $n<2$ is a simple but not safe criteria for pseudobulges (Fisher 2006), which is consistent with the observations of our sample.
Morphology and kinetic property are more reliable indicators than S\'ersic index.  In general, if the ``bulge" contains a nuclear bar, nuclear spirals, and/or nuclear ring, the bulge is identified as a pseudobulge. 
In $(V_{\rm m}/\sigma_{\rm b})$-$\epsilon$ diagram, pseudobulges locate above the oblate rotator line (e.g. Binney 1978), where $V_{\rm m}$ is the maximum line-of-sight rotational velocity of the bulge, $\sigma_{\rm b}$ is the rms stellar velocity dispersion of the bulge, and $\epsilon$ is the characteristic ellipticity in the region interior to the radius of $V_{\rm m}$. 
Bulges of different types (especially the boxy/peanut and discy bulges) often coexist in the same galaxy (Athanassoula 2005), the pseudobulges are found in all types (S0-Sm) of disk galaxies (Laurikainen et al. 2007).
The bulge properties of the disk galaxies in our sample are summarized in Table 3.

NGC 1068 has nuclear spirals, it is identified as pseudobulges using {\it HST} optical images by Drory \& Fisher (2007). 


NGC 2787 has a spectacular tilted nuclear dust ring of radius $\sim$1.$''$5-10$''$ (Sarzi et al. 2001; Erwin et al. 2003), a inner disk (Erwin 2003), and a bulge with disk-like kinetic property (Erwin 2007).
NGC 3384 has a inner disk (Erwin 2004) and a bulge with disk-like kinetic property (de Zeeuw et al. 2002). NGC 2787 and NGC 3384 are examples of the galaxies contain composite structures consisting of both pseudobulges and small inner classical bulges (Erwin 2007).

NGC 3079 has a peanut-shape bulge (L\"utticke, Dettmar, \& Pohlen 2000), which shows the cylindrical rotation (Shaw, Wilkinson \& Carter 1993), a character of disk-like pseudobulge.

NGC 3227 has a asymmetric molecular nuclear ring with a diameter of $\sim3''$ (Schinnerer, Eckart, \& Tacconi 2000), the S\'ersic index ($n=1.1$) of the bulge is also disk-like (Gadotti 2008).

NGC 3393 has nuclear spirals (Martini et al. 2003) and a possible nuclear bar (Jungwiert et al. 1997), which identify the pseudobulge.



The inner region of Circinus galaxy is bluer than typical bulge, that indicate young stellar populations (Wilson et al. 2000). The existence of a nuclear ring in CO emission (Curran et al. 1998) and near-UV light (Mu\~{n}oz Mar\'{\i}n et al. 2007) and a nuclear gas bar (Maiolino et al. 2000) identify the pseudobulge. 

IC 2560 has a boxy bulge  (L\"utticke, Dettmar, \& Pohlen 2000). The red-shifted features in the H$_2$O maser emission are slightly stronger than the blue-shifted, which is a typical characteristic seen in most megamasers and may be explained by a spiral structure of the maser-emitting nuclear disk (Ishihara et al. 2001).

Milky way has a bar-related boxy bulge (e.g. L\'opez-Corredoira et al. 2005), both the rotation (e.g. Minniti et al. 1992) and flattening (e.g. Sharples et al. 1990) of the bulge suggest a pseudobulge.

Most of the pseudobulges in our sample dwell in the barred disk galaxies (8/9), and most of our sample disk galaxies contain classical bulge are unbarred (11/14). Large scale bars are obviously predominant component in disk galaxies with pseudobugles, and serve as a strong indicator for the pseudobugles. In the end, we caution that the identification of pseudobulges is still somewhat uncertain, some galaxies need further discrimination. 

\begin{table}
\centering
\begin{minipage}{83mm}
  \caption{Bulge properties.}
  \begin{tabular}{llccccl}
  \hline
Galaxy & Type & $i$ ($^\circ$) & $n$ & $V_{\rm m}/\sigma_{\rm b}$ & $\epsilon$ & Structures  \\
 (1) &  (2)   &  (3)  & (4) & (5) & (6) & (7) 
   \\
  \hline
N224 & Sb & 78 & 2.2 & ... & ... & EB 
\\
N1023 & SB0 & 77 & 2.0 & 0.35 & 0.33 & EB 
\\
N3031 & Sab & 62 & 2.0 &  ... & ... & 
\\
N3115 & S0 & 82 & 4.4 & ... & ... & BB 
\\
N3245 & S0 & 67 & 4.0 & ... & ... & BB 
\\
N3414 & S0 & 77 & 3.8 & 0.09 & 0.21 &
\\
N3998 & S0 & 42 & 4.1 & ... & ... &
\\
N4151 & Sa & 21 & 3.0 &  ... & ... & 
\\
N4258 & SBbc & 71 & 3.0 &  ... & ... &
\\
N4459 & S0 &  47 & 3.5 &  ... & ... &
\\
N4596 & SB0 & 37 & 1.4 & 0.30 & 0.15 &
\\
N5128 & S0 & 48 & 4.0 &  ... & ... &
\\
N5252 & S0 & 67 & 6.0 &  ... & ... &
\\
N7457 & S0 & 76 & 2.0 & 0.62 & 0.44 & BB
\\
\hline
N1068 & SBb & 33 & 1.7 &  ... & ... & NS
\\
N2787 & SB0 & 66 & 1.0 & 0.91 & 0.35 & NR, ID
\\
N3079 & SBcd & 86 & ... & ... & ... & PB, NR
\\
N3227 & SBa & 69 & 1.1 & ... & ... & NR
\\
N3384 & SB0 & 59 & 1.0 & 0.77 & 0.29 & ID
\\
N3393 & SBa & 31 & ... & ... & ... & NS, NB?
\\
Circinus & Sb & 63 & ... & ... & ... & NR, NB
\\
IC 2560 & SBb & 66 & ... &  ... & ... & BB, NS?
\\
MW & SBbc & 90 & 1.3 &  ... & ... & BB, ID
\\
\hline
\end{tabular}
\\
Notes: (1) name of the galaxy. (2) Hubble type. (3) inclination of the disk are from Hyperleda or Ho 2007, except for 
N3384 (Neistein et al. 1999).
(4) S\'ersic index of the bulge. (5) $V_{\rm m}/\sigma_{\rm b}$ ratio within $R_{\rm e}$. (6) ellipticity of the bulge. 
(7) EB = elliptical bulge, BB = boxy bulge, PB = peanut bulge (L\"utticke, Dettmar, \& Pohlen 2000); NR = nuclear ring; NB = nuclear bar; NS = nuclear spirals; ID = inner disk.
\end{minipage}
\end{table}

\subsection{Comments on individual objects}

{\it Early-type bulges}:

NGC 224 (M31), NGC 3031 (M81) and NGC 3115 in our sample are typical classical bulges listed in KK04. 

NGC 221 (M32).---It is a small, low-luminosity elliptical, has a main body of S\'ersic profile with index $n<4$ and extra light near the center above the inward extrapolation of the outer S\'ersic function (Kormendy 1999).

NGC 1399.---The black hole mass ($M_{\rm bh}=12^{+5}_{-6}\times10^8$ M$_\odot$) is taken from Houghton et al. (2006), which is consistent with the result of Gebhardt et al. (2007), $M_{\rm bh}=(5.1\pm0.7)\times10^8$ M$_\odot$. 

NGC 3115 and NGC 3245 have boxy bulges, while their bulge S\'ersic index is $n\simeq4$. We suggest central classical bulges and outer boxy bulges coexist in these galaxies.

NGC 3377.---We estimate the 1$\sigma$ black hole mass error according to the Fig. 12 of Copin et al. (2004).

NGC 3379.---The black hole mass is measure by stellar kinematics ($M_{\rm bh}=1.0^{+1.0}_{-0.4}\times10^8$ M$_\odot$; Gebhardt et al. 2000b) and gas kinematics ($M_{\rm bh}=1.4^{+0.3}_{-0.8}\times10^8$ M$_\odot$; Shapiro et al. 2006), we take the average value of $M_{\rm bh}=(1.2\pm0.6)\times10^8$ M$_\odot$. 

NGC 3998.---The black hole mass is linearly modified with distance taken from Tonry et al. (2001), the 1$\sigma$ error is estimated from Fig. 12 of De Francesco et al. (2006).

NGC 4151.---The black hole mass $M_{\rm bh}=3^{+1}_{-2}\times10^7$ M$_\odot$ is a mean value of the measurement using stellar kinematics ($\sim$4$\times10^7$ M$_\odot$, Onken et al. 2007) and gas kinematics ($3.2^{+0.8}_{-2.3}\times10^7$ M$_\odot$ [corrected for distance], Hicks \& Malkan 2008), which is consistent with the RM measurement ($4.57^{+0.57}_{-0.47}\times10^7$ M$_\odot$, Bentz et al. 2006). The stellar velocity dispersion is taken from Nelson et al. (2004) measured with an aperture of $6''.5$.

NGC 4486A.---We estimate the 1$\sigma$ black hole mass error and velocity dispersion according to the Fig. 6 and Fig. 4 of Nowak et al. (2007).

NGC 4596 has a relatively exponential bulge with S\'erisic index $n=1.4$ (Laurikainan et al. 2005), 
but the bulge kinematics (Bettoni \& Galletta 1997) is ellipsoidal.

NGC 4649.---We estimate the $\sigma_{\rm e}$ using the stellar velocity dispersion profile measured by Bridges et al. (2006).

NGC 5128.---The distance ($D=3.5$ Mpc) and black hole mass ($M_{\rm bh}=4.5^{+0.6}_{-0.4}\times10^7$ M$_\odot$) is taken from Neumayer et al. (2007), we estimate the 1$\sigma$ black hole mass error according to their Fig. 12. This black hole mass is consistent with the result of H\"aring-Neumayer et al. (2006), $M_{\rm bh}=6.1^{+0.6}_{-0.8}\times10^7$ M$_\odot$.

NGC 5252.---The stellar velocity dispersion is taken from Nelson \& Whittle (1995) measured with an aperture of $1''.5\times2''.3$.

NGC 7457.---This galaxy has a young box-shape bulge (L\"utticke, Dettmar, \& Pohlen 2000; Kormendy \& Kennicutt 2004). Kormendy (1993) suggests it as a pseudobulge according to its small $\sigma_*$. But the kinetic property (Cappellari et al. 2007a) identifies it as a classical bulge.

The black hole masses of NGC 2974, NGC 3414, NGC 4552, NGC 4621, NGC 5813 and NGC 5846 are taken from Cappellari et al. (2007b) with a formal error of 10\% (1$\sigma$ C.L.), which is approximately converted from the 30\% error at 3$\sigma$ C.L.

{\it Pseudobulges}:

NGC 1068.---A prominent bar is present in the 2MASS infrared image (Men\'endez-Delmestre et al. 2007), while the NED (RC3) type is Sb. The stellar velocity dispersion is taken from Nelson \& Whittle (1995) measured with an aperture of $1''.5\times2''.3$.

NGC 2787.---T02 gives $\sigma_{\rm e}=140$ km s$^{-1}$, while in the literature much higher value are quoted, e.g., 210 km s$^{-1}$ (Dalle Ore et al. 1991), 205 km s$^{-1}$ (Neistein et al. 1999), 202 km s$^{-1}$ (Barth et al. 2002), and 257 km s$^{-1}$ (Erwin et al. 2003). As pointed by Aller et al. (2007), this discrepancy may stem from the central peak in the velocity dispersion profile. We estimate $\sigma_{\rm e}$ and $\sigma_{\rm c}$ from Fig. 1 of Bertola et al. (1995).

NGC 3079.---The accretion disk of the SMBH is probably thick, flared, and clumpy, in contrast to the compact, thin, warped, differentially rotating disk in the archetypal maser galaxy NGC 4258 (Kondratko, Greenhill, \& Moran 2005). The mass of SMBH thus may be underestimated. The stellar velocity dispersion is taken as the central value from Shaw et al. (1993).

NGC 3227.---The mass $2.0^{+1.4}_{-1.3}\times10^7$ M$_\odot$ is a mean value of the measurement using stellar kinematics ([0.7-2.0]$\times10^7$ M$_\odot$, Davies et al. 2006) and gas kinematics ($2.3^{+1.1}_{-0.5}\times10^7$ M$_\odot$ [corrected for distance], Hicks \& Malkan 2008), which is consistent with the RM measurement ($[4.2\pm2.1]\times10^7$ M$_\odot$, Peterson et al. 2004). The stellar velocity dispersion is estimated by the author from the Fig. 6, Fig. 11 and Fig. 16 of Davies et al. (2006).

NGC 3393.---The stellar velocity dispersion is taken from Terlevich et al. (1990) measured with an aperture of $2''.1$.

Circinus.---The stellar velocity dispersion is taken as mean value of central $10''$ from Fig. 6 of Maiolino et al. (1998). The distance is taken from Karachentsev et al. (2007), which is different from the value in $M_{\rm bh}$ measurement (Greenhill et al. 2003).

IC 2560.---The stellar velocity dispersion is measured with an aperture of $2''\times2''.46$ (Garcia-Rissmann et al. 2005).


We do not include the following galaxies in our sample:

NGC 1300 and NGC 2748.---NGC 1300 has spiral-like dust lanes down to the center, wrapping around the nucleus (Scarlata et al. 2004), and connecting to the leading edge of a strong kpc-scale bar (Martini et al. 2003), which suggests gas inflow in both large and small scale. The bulge of NGC 1300 is disk-like with S\'ersic index $n=1.3$ (Laurikainen et al. 2004), and identified as a pseudobulge by Kormendy \& Fisher (2005). 
The nuclei of N2748 is not well defined in {\it HST} observations due to dust lane obscuration (Scarlata et al. 2004). 
The masses of SMBHs in NGC 1300 ($6.6^{+6.3}_{-3.2}\times10^7\ {\rm M}_\odot$) 
and NGC 2748 ($4.4^{+3.5}_{-3.6}\times10^7\ {\rm M}_\odot$) are measured by nuclear gas kinematics (Atkinson et al. 2005). 
The authors neglected the effect of extinction by dust in the analysis of the gaseous rotation curves. 
The extinction in the H band which was used to measure the black hole masses yield an underestimation in luminosity of $\sim20\%$ for NGC 1300 and $\sim58\%$ for NGC 2748 in their central 50 pc regions. Furthermore, the extinction is probably uneven due to the complex dust structures. Therefore, the $M_{\rm bh}$ in two galaxies are not well estimated. 
Their results are much higher than the prediction of  $M_{\rm bh}$-$\sigma_*$ or $M_{\rm bh}$-$L$ relations 
($7\times10^6\ {\rm M}_\odot$ or $2\times10^7\ {\rm M}_\odot$ for NGC 1300; 
$6\times10^6\ {\rm M}_\odot$ or $7\times10^6\ {\rm M}_\odot$ for NGC 2748). 
In fact, the {\it HST} can only barely resolved the sphere of influence of SMBHs with the mass predicted by the $M_{\rm bh}$-$\sigma_*$ or $M_{\rm bh}$-$L$ relations ($r_{\rm bh}=0''.04\sim0''.1$ for NGC 1300; $r_{\rm bh}=0''.03$ for NGC 2748).

NGC 2778 and NGC 4303.---{\it HST} cannot resolve the influence radius of the SMBHs in NGC 2778 ($2r_{\rm bh}/r_{\rm res}=0.8$, Gebhardt et al. 2003) and in NGC 4303 ($2r_{\rm bh}/r_{\rm res}=0.9$, Pastorini et al. 2007). The zero black hole mass model for NGC 2778 is ruled out at only the 95\% C.L. (Gebhardt et al. 2003).

NGC 4342.---The mass of the central SMBH measured by Cretton \& van den Bosch (1999) make this object a outlier for the $M_{\rm bh}$-$L$ relation (Marconi \& Hunt 2003) and the $M_{\rm bh}$-$M_{\rm b}$ relation (H\"aring \& Rix 2004). According to the analysis of Valluri, Merritt \& Emsellem (2004), no best-fit value of $M_{\rm bh}$ can be found for this object, the published value $3.0^{+1.7}_{-1.0}\times10^8$ M$_\odot$ is probably overestimated.

NGC 4350 (Pignatelli et al. 2001), NGC 4486B (Kormendy et al. 1997), and NGC 4594 (Kormendy et al. 1996).---Their mass measurement is uncertain due to lack of three-integral stellar dynamical models. 

NGC 4374.---As pointed by T02, the published measurement of $M_{\rm bh}$ in NGC 4374 (M84) is differ from each other beyond the stated errors.


NGC 7582.---The influence radius of the SMBH is not well resolved, $2r_{\rm bh}/r_{\rm res}=0.7$ (Wold et al. 2006).

The more reliable and accurate mass measurement of these SMBHs in the future will contribute to the poorly sampled low-mass end of the  $M_{\rm bh}$-$\sigma_*$ relation, especially for the pseudobulges. 

\section{The estimation of $M_{\rm bh}$-$\sigma_*$ relations}


\begin{figure*}
\centerline{\includegraphics[width=8.5cm]{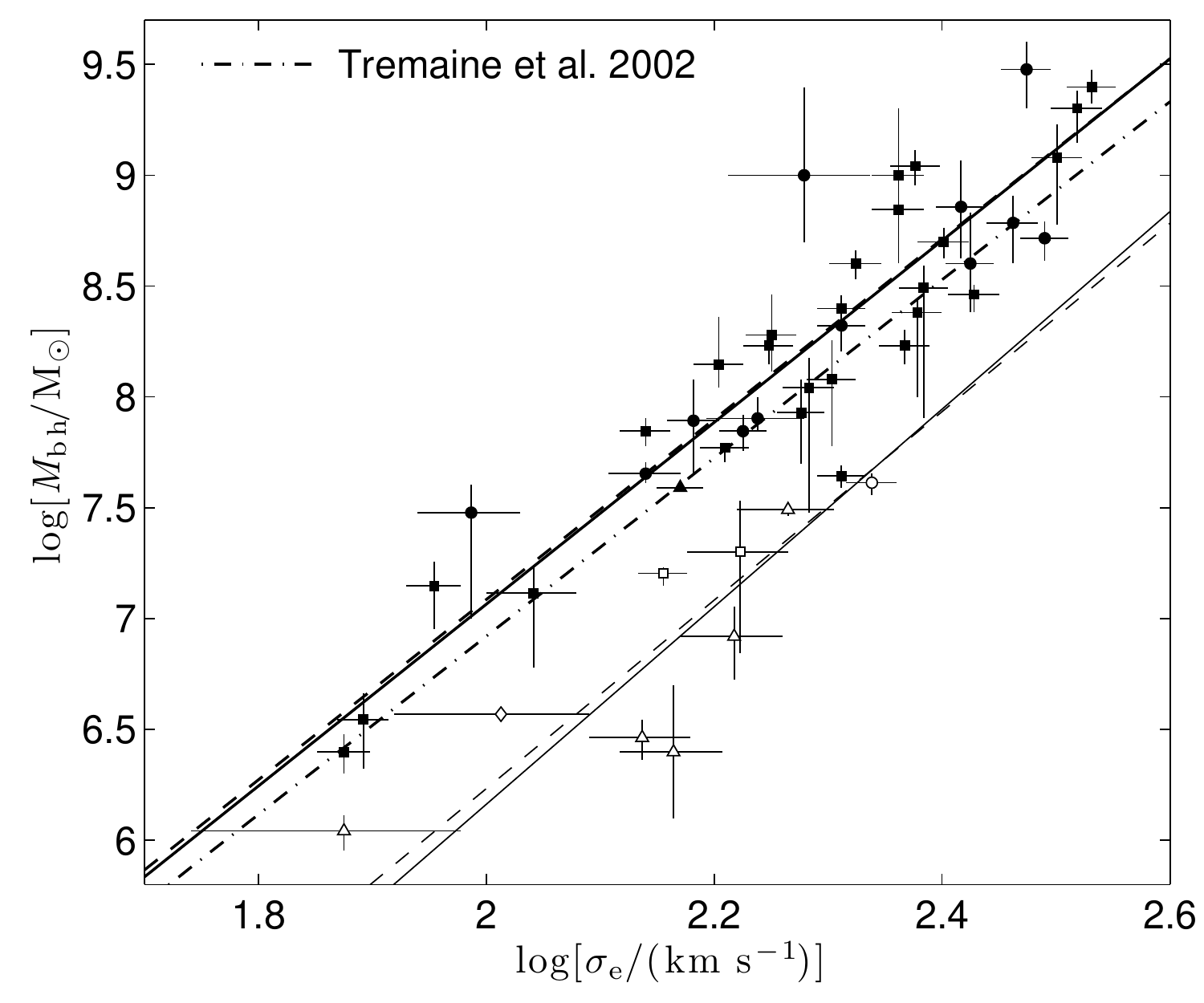}\includegraphics[width=8.5cm]{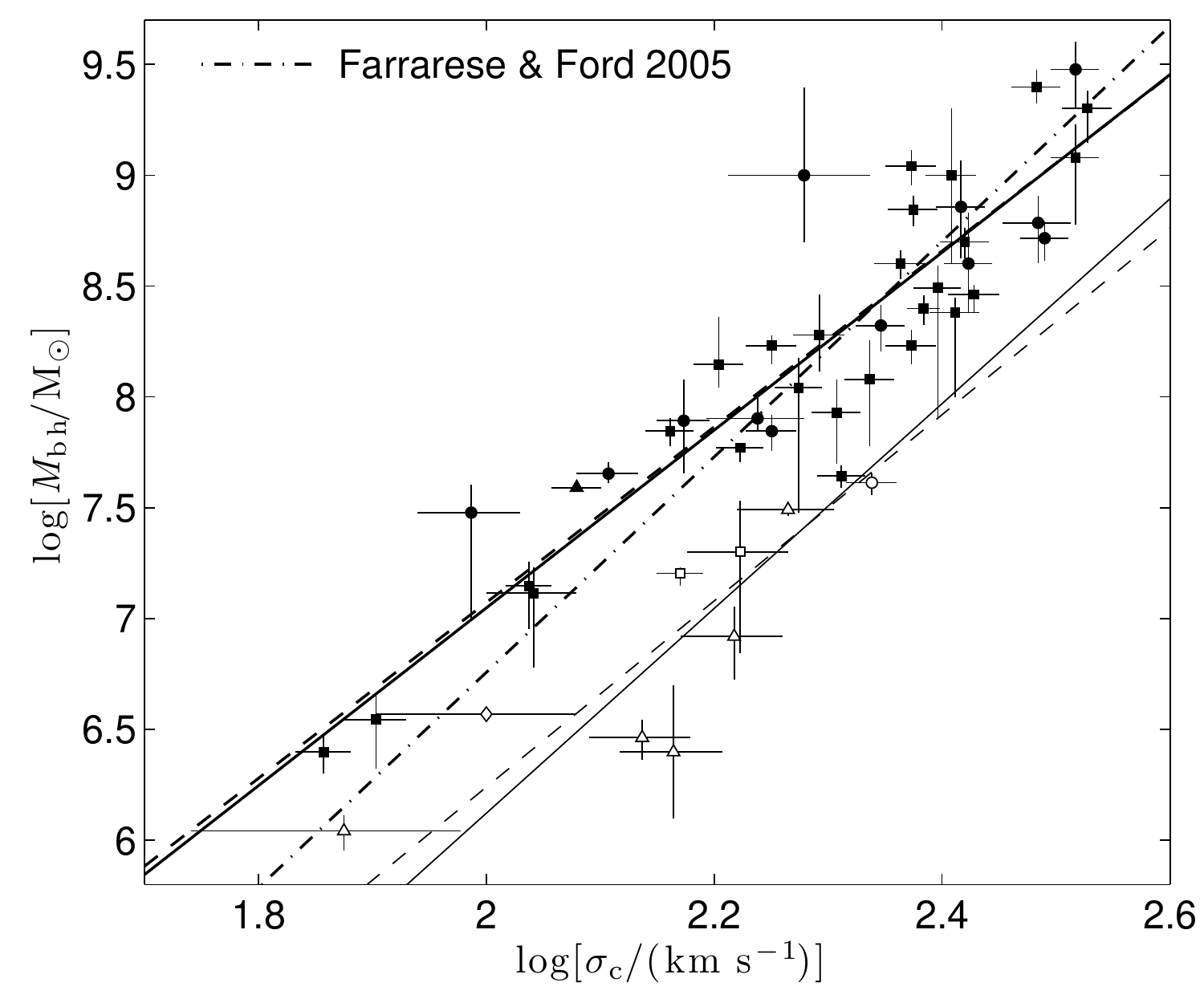}} 
\centerline{\includegraphics[width=8.5cm]{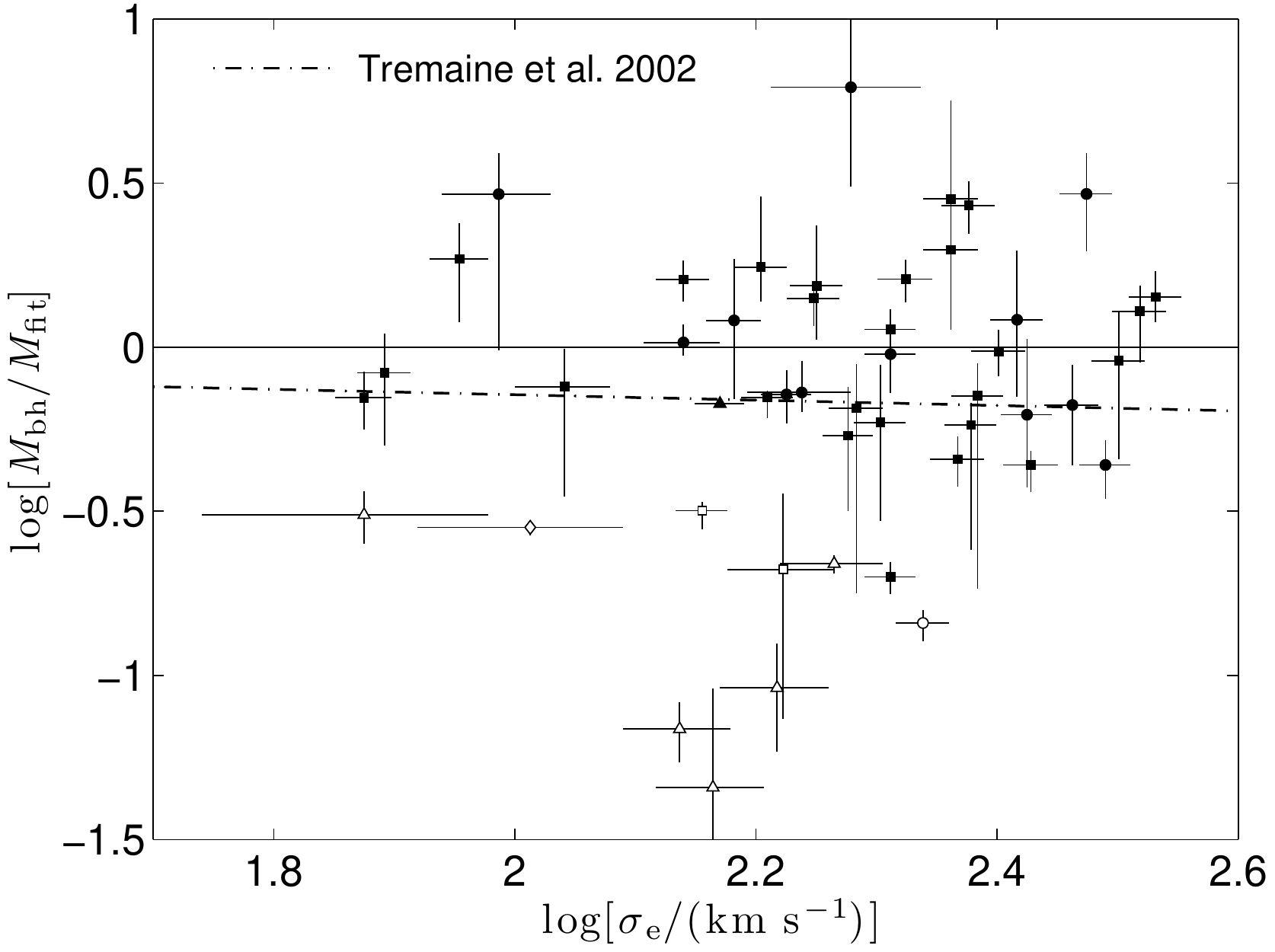}\includegraphics[width=8.5cm]{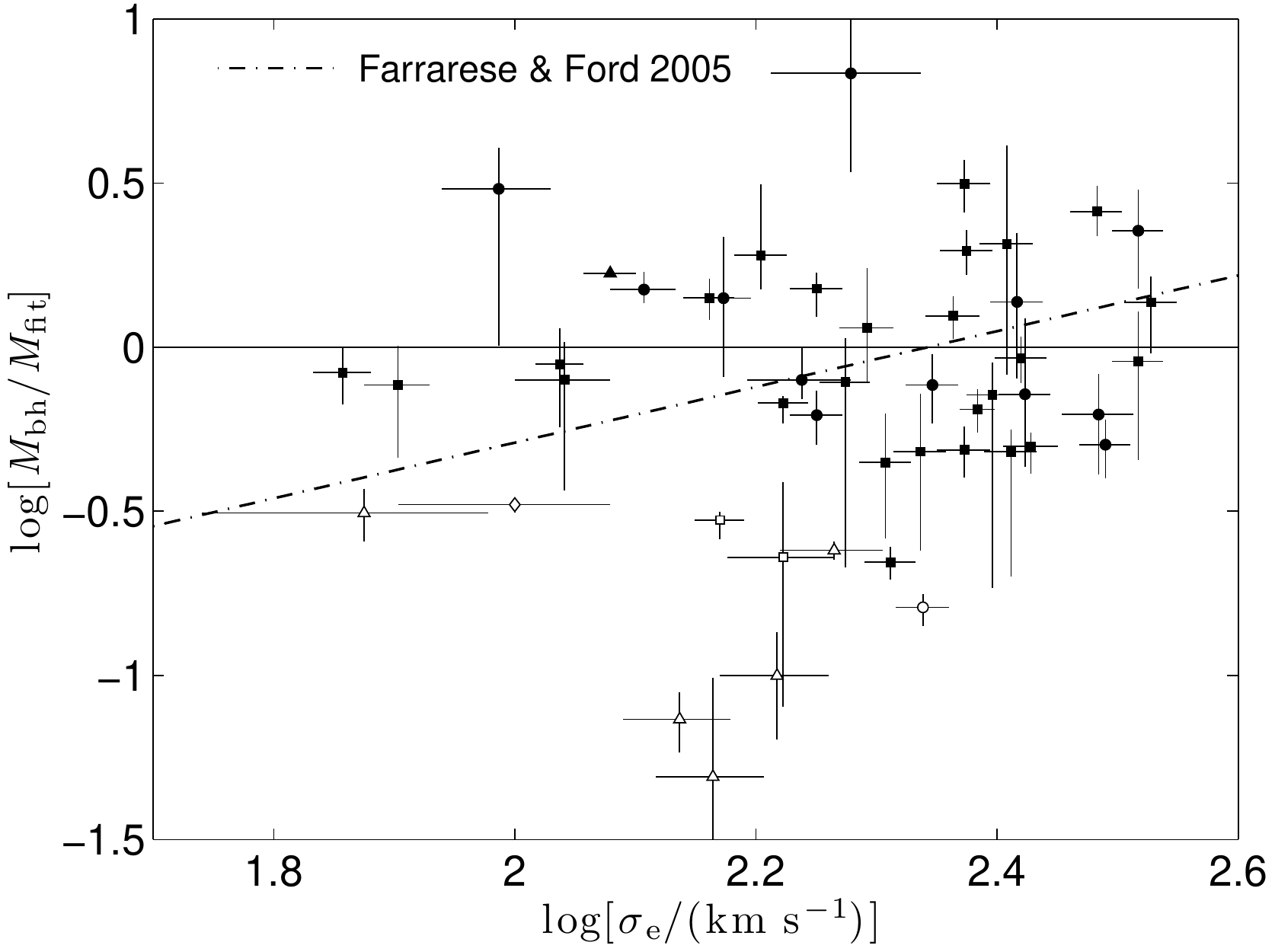}} 
\centerline{\includegraphics[width=8.5cm]{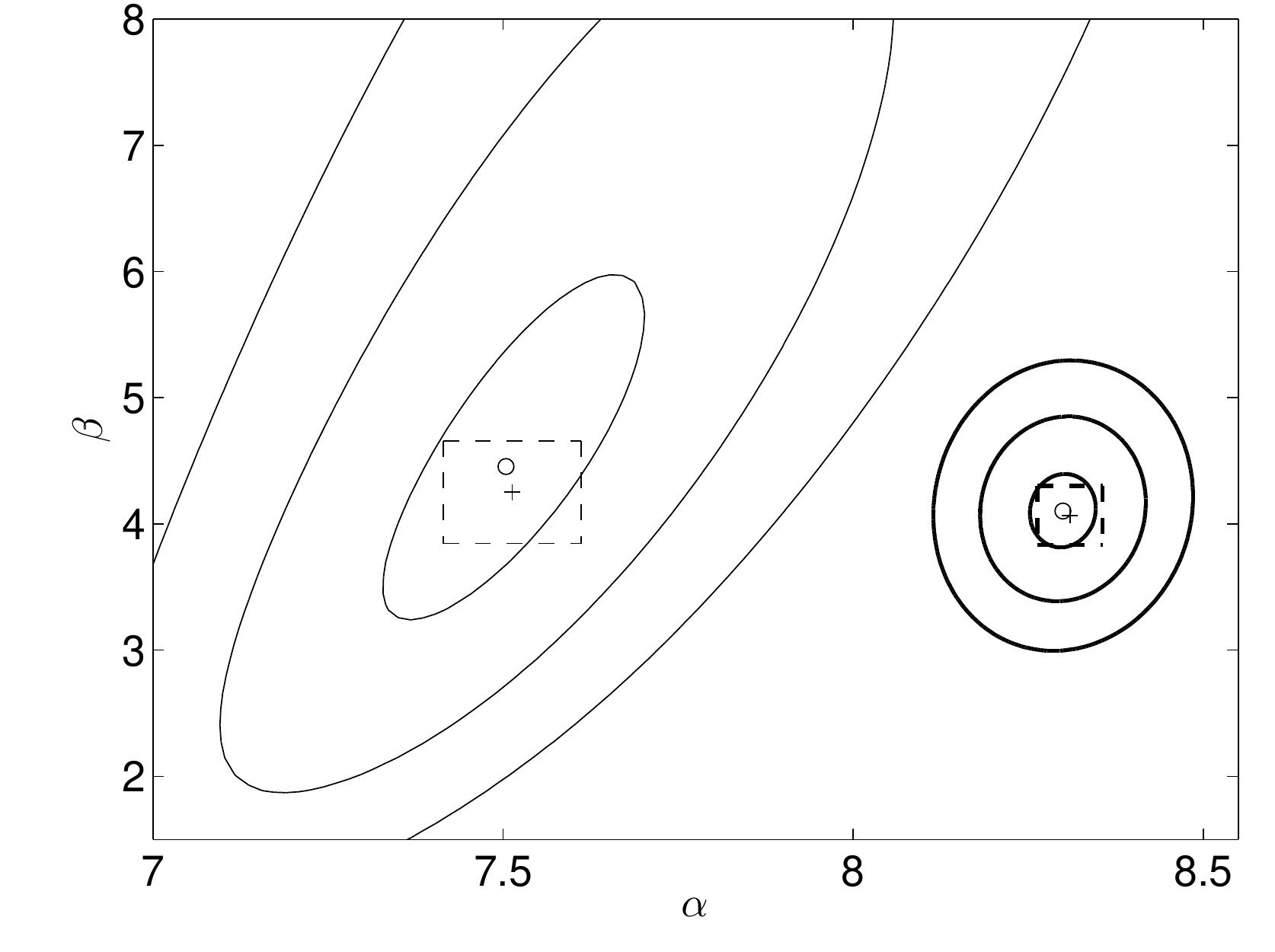}\includegraphics[width=8.5cm]{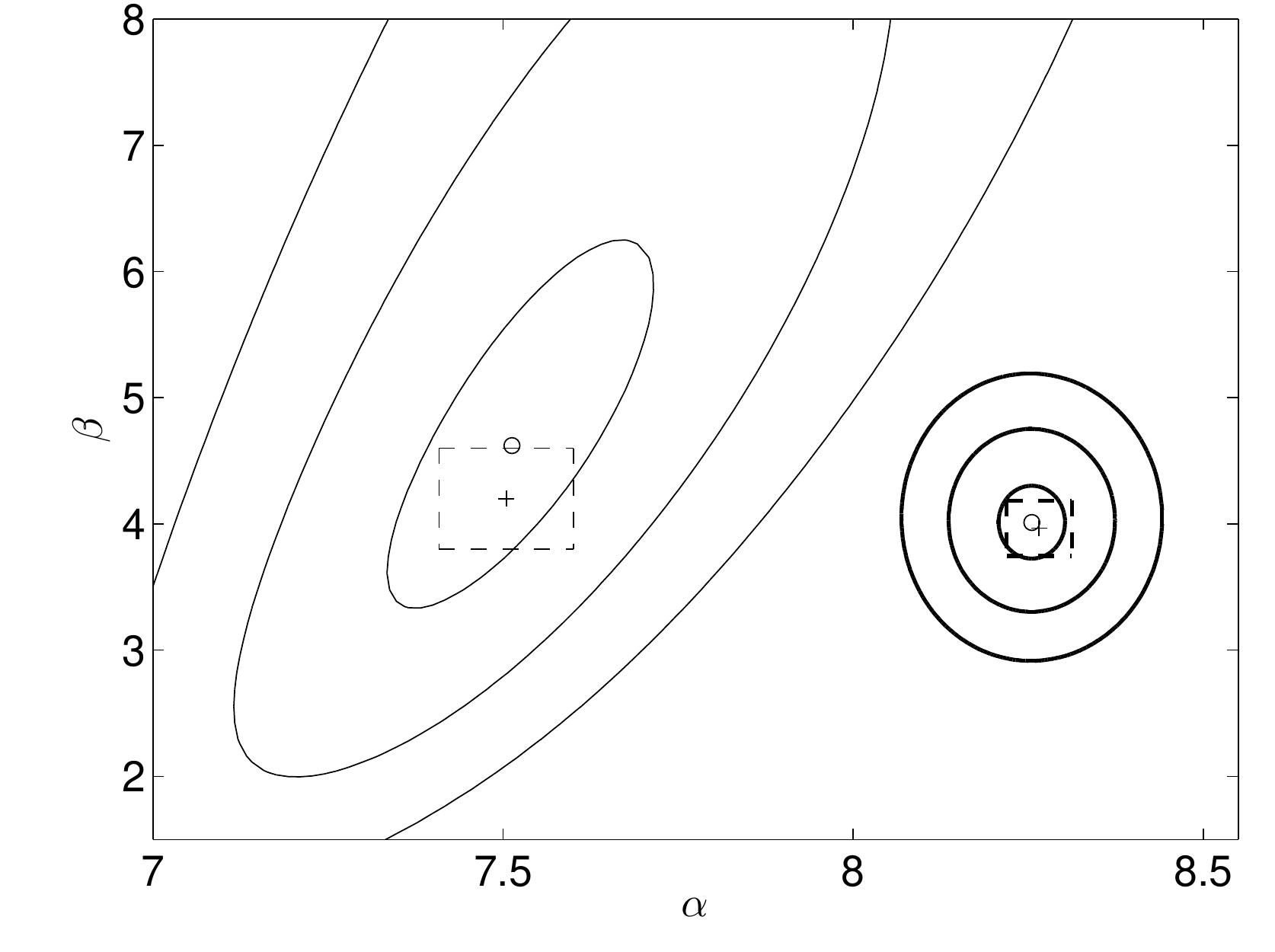}} 
\caption{{\it Left-top}: The  $M_{\rm bh}$-$\sigma_{\rm e}$ relations for early-type bulges (filled markers, thick lines) and pseudobulges (open markers, thin lines). Mass measurement based on stellar kinematics are denoted by squares, on gas kinematics by circles, on masers by triangles, and on stellar proper motions by diamond. The solid lines are fitting results using $\chi^2$ estimator, dashed lines using AB estimator. {\it Left-middle}: Residuals between the measured $M_{\rm bh}$ and the prediction by best-fit relation using $\chi^2$ estimator for early-type bulges. The markers are the same as the top panel. The best-fit relation of T02 is shown for comparison. {\it Left-bottom}: The contours are 1$\sigma$, 2$\sigma$ and 3$\sigma$ limits of $\chi^2$ fitting of $\alpha$ and $\beta$. The central circles denote best-fit value. The dashed squares and their central cross markers denote the best-fit value and uncertainties using AB estimator. The thick and thin lines are results of early-type bulges and pseudobulges respectively. {\it Right}: Similar as the left panel, but for the  $M_{\rm bh}$-$\sigma_{\rm c}$ relations, the best-fit of T02 is substituted by one of FF05.}
\end{figure*}

\begin{table*}
\centering
\begin{minipage}{180mm}
  \caption{The fitting results of the $M_{\rm bh}$-$\sigma_*$ relation.}
  \begin{tabular}{cccccccccccc}
  \hline
&\multicolumn{5}{c}{$M_{\rm bh}$-$\sigma_{\rm e}$} && \multicolumn{5}{c}{$M_{\rm bh}$-$\sigma_{\rm c}$}\\
\cline{2-6} \cline{8-12} 
Host type  & $\alpha_{\chi^2}$  & $\beta_{\chi^2}$  &  $\epsilon_0$  & $\alpha_{\rm AB}$  &  $\beta_{\rm AB}$  &
                & $\alpha_{\chi^2}$  & $\beta_{\chi^2}$  &  $\epsilon_0$  & $\alpha_{\rm AB}$  &  $\beta_{\rm AB}$ 
\\
\hline
early-type bulges & 8.30$\pm$0.05 &4.10$\pm$0.28 & 0.26 &8.31$\pm$0.05&4.07$\pm$0.24 &
                        & 8.26$\pm$0.05 &4.01$\pm$0.28 & 0.27 &8.27$\pm$0.05 &3.97$\pm$0.22
\\
pseudobulges & 7.50$\pm$0.18 &4.46$\pm$1.30 & 0.25 &7.51$\pm$0.10&4.25$\pm$0.41 &
                     & 7.51$\pm$0.17 & 4.62$\pm$1.32 & 0.24 &7.50$\pm$0.10&4.20$\pm$0.40
\\
all galaxies & 8.18$\pm$0.06 &4.57$\pm$0.37 & 0.42 &8.18$\pm$0.06&4.50$\pm$0.30 &
                 & 8.14$\pm$0.06 &4.52$\pm$0.36 & 0.39 &8.14$\pm$0.06&4.59$\pm$0.32
\\
core ellipticals & 8.28$\pm$0.11 &4.84$\pm$0.78 & 0.24 &8.27$\pm$0.16&5.62$\pm$1.47 &
                      & 8.17$\pm$0.11 &5.33$\pm$0.81 & 0.21 &8.15$\pm$0.21& 6.25$\pm$1.79
\\
normal ellipticals & 8.28$\pm$0.07 &3.97$\pm$0.38 & 0.29 &8.28$\pm$0.07&4.00$\pm$0.36 &
                           & 8.22$\pm$0.06 &3.82$\pm$0.36 & 0.29 &8.23$\pm$0.07&3.81$\pm$0.31
\\
\hline
\end{tabular}
\end{minipage}
\end{table*}

\subsection{The linear fitting algorithm}

Different authors use different algorithms to estimate the log-linear $M_{\rm bh}$-$\sigma_*$ relation eq. (1).
G00 and T02 fit the relation by minimizing ``$\chi^2$ estimator" (Press et al. 1992)
\beq
\chi^2=\sum_{i=1}^{N}\frac{(y_i-\alpha-\beta x_i)^2}{\epsilon^2_{yi}+\beta^2\epsilon^2_{xi}},
\eeq
where $x=\log(\sigma/200\ {\rm km\ s^{-1}})$, $y=\log(M_{\rm bh}/{\rm M}_\odot)$, 
$\epsilon_{xi}$ and $\epsilon_{yi}$ are errors of $x$ and $y$ measurements.  
In order to account for the intrinsic scatter in the $M_{\rm bh}$-$\sigma_*$, T02 replace $\epsilon_{yi}$ by 
$(\epsilon^2_{yi}+\epsilon^2_0)^{1/2}$, where $\epsilon_0$ is a constant denotes the intrinsic scatter of $y$ at a fixed $x$. 
The value of $\epsilon_0$ is set so that the best-fit reduced $\chi^2$ is equal to its expectation value of unity, T02 find $\epsilon_0=0.27$ for their sample. 
The $1\sigma,\ 2\sigma,\ 3\sigma$ uncertainties in $\alpha$ and $\beta$ are given by the upper and lower limits of $\alpha$ and $\beta$ for which $\chi^2-\chi^2_{\rm min}=1.01,\ 6.75,\ 16.68$.

FF05 use the ``AB estimator" described by Akritas \& Bershady (1996) to determine the best-fit 
$\alpha$, $\beta$ and their uncertainties. Both $\chi^2$ and AB estimators have the advantages of accounting for measurement uncertainties in both variables and are asymptotically normal and consistent. In our analysis, the AB estimator gives the similar estimate for the slope $\beta$ (cf. Figure 2, Figure 3, and Table 4).
A review of the comparison of the two estimator can be found in T02. 

In the subsequent analysis, we use both logarithms to estimate $\alpha$ and $\beta$, and denote them as $\alpha_{\chi^2}$, $\beta_{\chi^2}$ and $\alpha_{\rm AB}$, $\beta_{\rm AB}$, respectively. 

\subsection{The slope and normalization of the relations}

The fitting results of the $M_{\rm bh}$-$\sigma_{\rm e}$ relation and the $M_{\rm bh}$-$\sigma_{\rm c}$ relation are shown in Figure 2. The best-fit value and uncertainties of $\alpha$, $\beta$ and $\epsilon_0$ are summarized in Table 4. We notice the  $M_{\rm bh}$-$\sigma_{\rm e}$ and  $M_{\rm bh}$-$\sigma_{\rm c}$ relations are roughly the same, and the results derived by two fitting algorithms are very close, both within 1$\sigma$ uncertainty. 

The slope $\beta$ for the early-type bulges derived by two fitting algorithms and two dispersion definitions, is in the range of 3.9$\sim$4.1 with uncertainties of $\lesssim$0.28. It is consistent with the result of T02 ($\beta=4.02\pm0.32$).  
We take the mean fitting results by $\chi^2$ estimator as the ``uniform''  $M_{\rm bh}$-$\sigma_*$ relation for the early-type bulges:
\beq
\alpha=8.28\pm0.05,\ \ \ \beta=4.06\pm0.28.
\eeq
The case for the pseudobulges are similar, but with the different value. Their best-fit of $M_{\rm bh}$-$\sigma_*$ by $\chi^2$ estimator is 
\beq
\alpha=7.50\pm0.18,\ \ \ \beta=4.5\pm1.3.
\eeq
The intrinsic scatter for the early-type bulges and pseudobulges are 0.27 and 0.25 respectively. The $M_{\rm bh}$ measurement and the prediction by the best-fit relation are compared in the middle panels of Figure 2. The best-fits of T02 and FF05 are plotted in the same figures for comparison. Most of the pseudobulges lie below the prediction of relation for the early-type bulges and of all the three previous formula. 

The discrepancy between the $M_{\rm bh}$-$\sigma_*$ relations for the early-type bulges and pseudobulges is explicit in the bottom panels of Figure 2. They are distinct over 3$\sigma$ significance level. The Kolmogorov-Smirnov test gives a very low probability ($P<2\times10^{-8}$) that the $M_{\rm bh}$-$\sigma_*$ relation for pseudobulge agrees with the relation for the early-type bulges. 

The relation for the pseudobulges has a similar slope but $0.78\pm0.18$ dex smaller normalization than the relation for the early-type bulges. On average, the SMBHs in early-type bulges are as 6 times massive as ones in pseudobulges with the same velocity dispersion. T02's normalization ($\alpha=8.13\pm0.06$) is just between our estimates of the early-type bulges and pseudobulges, may be a result of fitting for a mixed sample of two types of bulges. 

We also fit the $M_{\rm bh}$-$\sigma_*$ relation for a combined sample of all the objects, obtain a steeper slope than that of the both samples. It is easy to be understood that, the $M_{\rm bh}$-$\sigma_*$ relation for pseudobulges is roughly parallel but lower than that for early-type bulges, the high mass end dominant early-type bulges and the low mass end dominant pseudobulges will thus determine a steeper relation for the combined sample. It can partially explain the steep slope derived by FF05 (cf. the right panels of Figure 2).

\begin{figure*}
\centerline{\includegraphics[width=8.5cm]{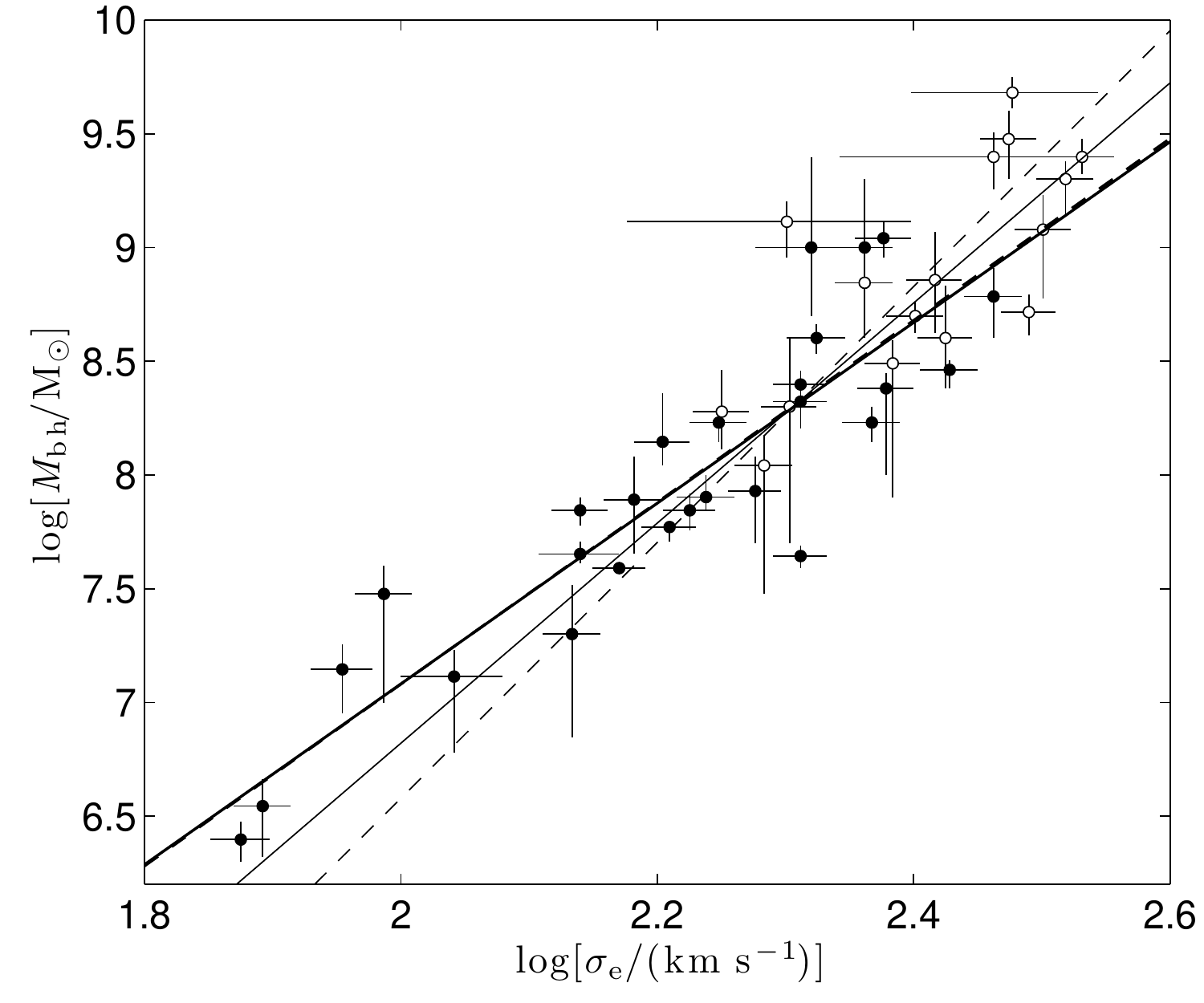}\includegraphics[width=8.5cm]{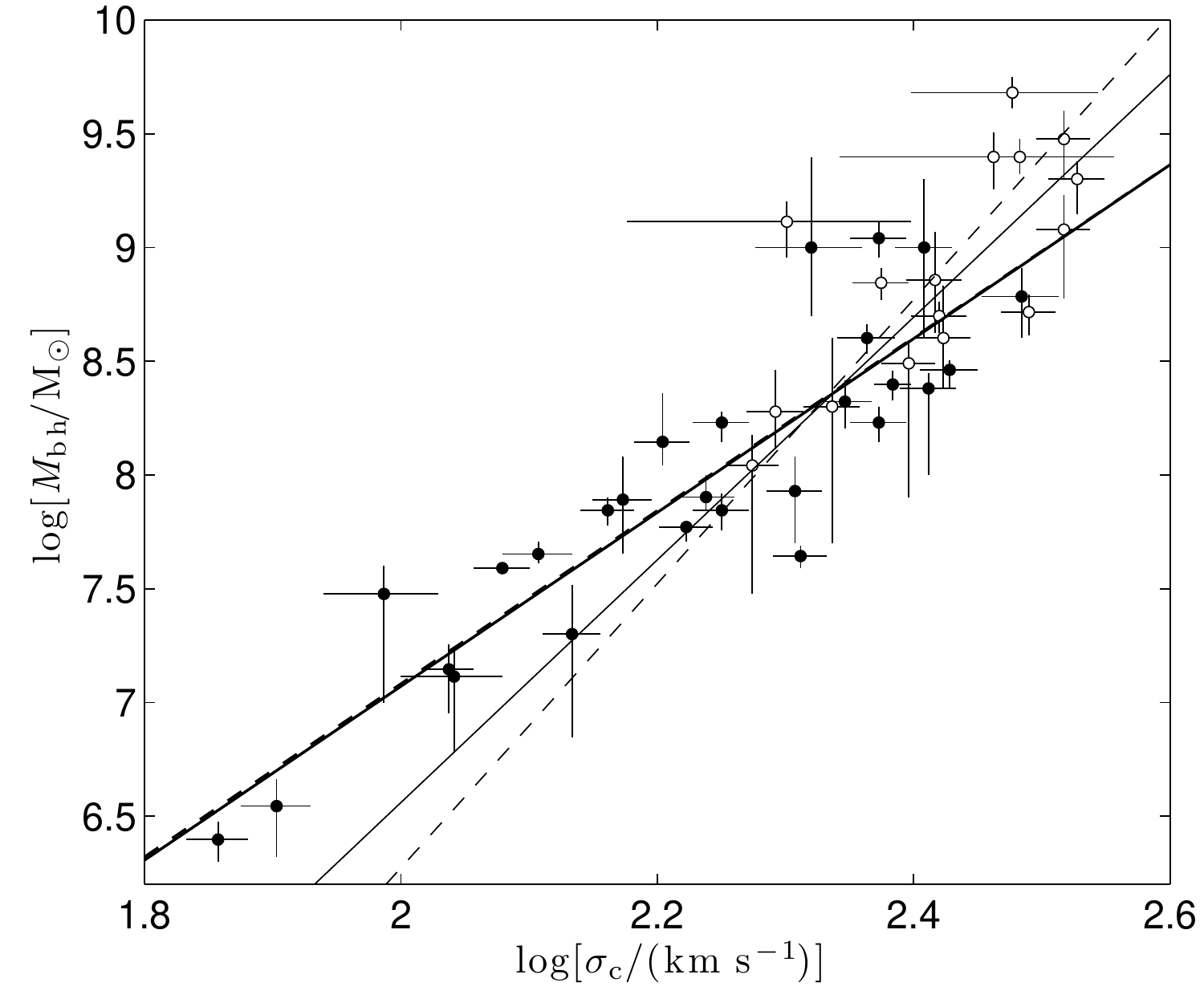}} 
\centerline{\includegraphics[width=8.5cm]{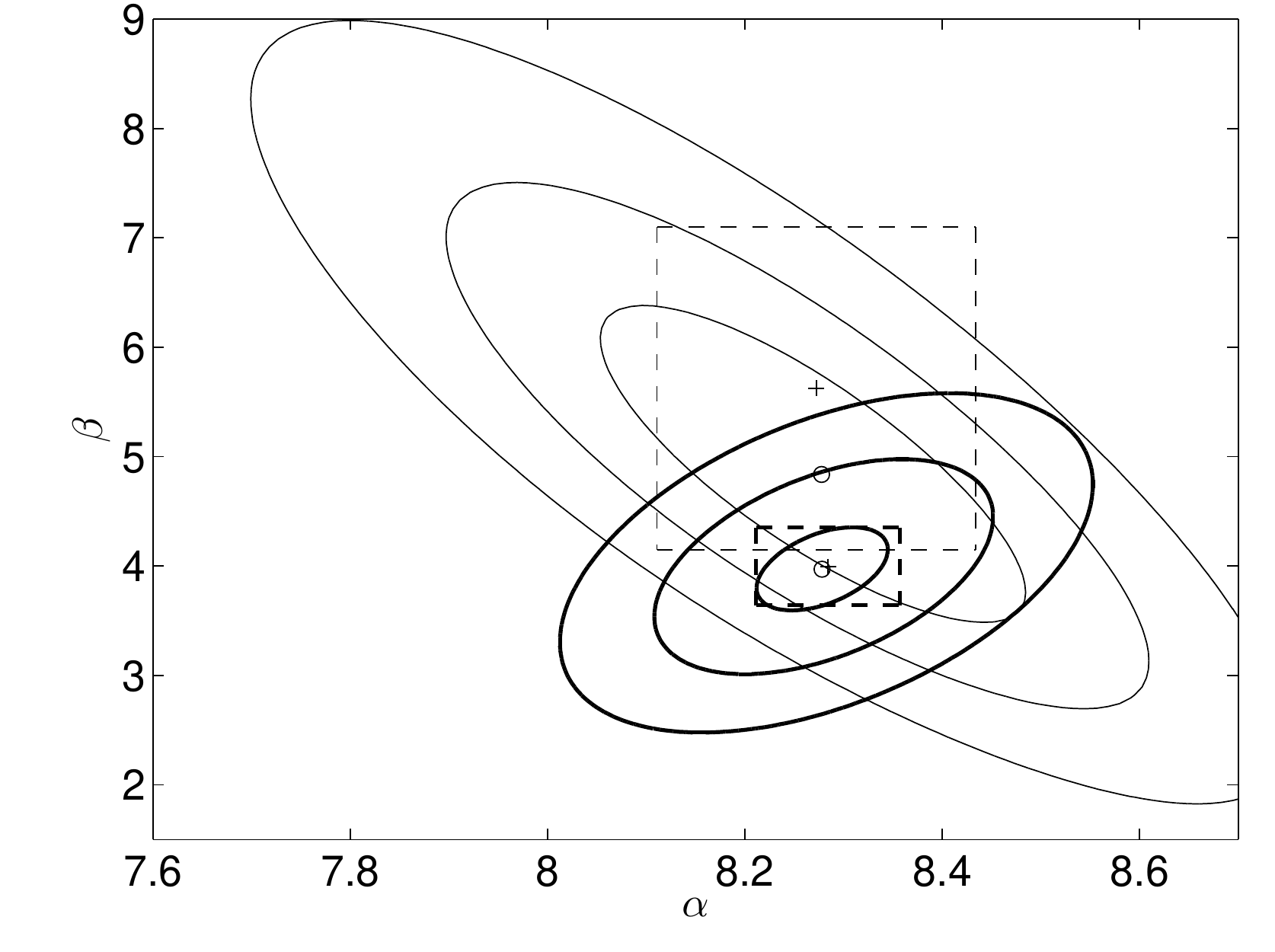}\includegraphics[width=8.5cm]{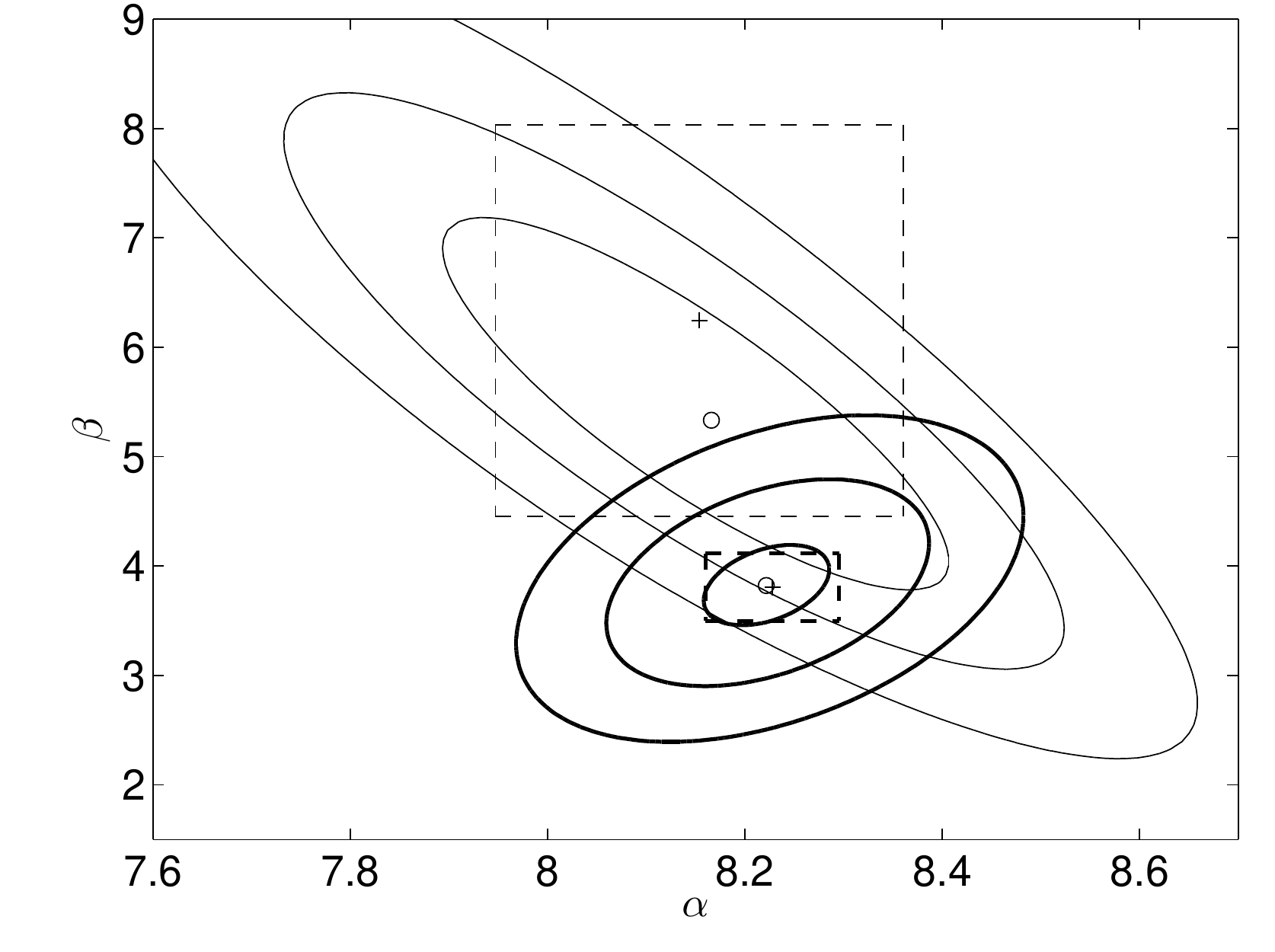}} 
\caption{{\it Left-top}: The best-fit  $M_{\rm bh}$-$\sigma_{\rm e}$ relations for ``core" galaxies (open circles, thin lines) and the other early-type galaxies (filled circles, thick lines). The solid lines and dashed lines denote fitting results using $\chi^2$ and AB estimator respectively. {\it Left-bottom}: The contours are 1$\sigma$, 2$\sigma$ and 3$\sigma$ limits of $\chi^2$ fitting of $\alpha$ and $\beta$ in eq. (1). The central circles denote best-fit value. The dashed squares and their central cross markers denote the best-fit value and uncertainties using AB estimator. The thin and thick lines are results of the ``core" elliptical galaxies and the normal elliptical galaxies. {\it Right}: Similar as the left panel, but for the  $M_{\rm bh}$-$\sigma_{\rm c}$ relations.}
\end{figure*}

\subsection{The $M_{\rm bh}$-$\sigma_*$ relation for the ``core'' elliptical galaxies}

Some elliptical galaxies are found to have cores in their central brightness profiles (Lauer et al. 1995). 
A core is evident as a radius at which the steep envelope of the galaxy ÔÔbreaksÕÕ and becomes a shallow profile in logarithmic coordinates.
The cores are believed formed by binary BHs in the dissipationless (``dry") mergers. 
These remnants of dry mergers will preserve the $M_{\rm bh}$-$L$ relation but deviate from the $M_{\rm bh}$-$\sigma_*$ relation, 
because the Faber-Jackson relationship $L\propto\sigma^7$ for the ``core'' galaxies are much steeper than than the classic $L\propto\sigma^4$ for the ``normal'' elliptical galaxies (Lauer et al. 2007a).
On the other hand, the SMBHs masses $M_{\rm bh}$ predicted by the $M_{\rm bh}$-$\sigma_*$ relation is systematically smaller the masses predicted by the $M_{\rm bh}$-$L$ relation for high-luminosity galaxies, such as brightest cluster galaxies (BCGs). The BCGs have large cores, which are probably formed in the dry mergers. Hence, the $M_{\rm bh}$-$\sigma_*$ relation will break at the high mass range due to the core galaxies made by dry mergers.

Inspired by these arguments, we attempt to compare the $M_{\rm bh}$-$\sigma_*$ relations of core elliptical galaxies in our sample and of the other normal early type galaxies. We augment the 13 core elliptical galaxies in Table 1 with the brightest group galaxy (BGG) Cyg A (Tadhunter et al. 2003; Thornton et al. 1999), and the BCGs in galaxy cluster Abell 1836 and Abell 3565 (Dalla Bont\`a et al. 2007). The BGGs and BCGs are believed to have similar physical origin with core elliptical galaxies. Their central black hole mass measured by gas kinematics and central stellar velocity dispersion\footnote{These objects are excluded in section 2 for their relatively uncertain $\sigma_*$.} are: Cyg A, $M_{\rm bh}=[2.5\pm0.7]\times10^9$ M$_\odot$, $\sigma_*$=290$\pm$70 km s$^{-1}$; Abell 1836 BCG, $M_{\rm bh}=4.8^{+0.8}_{-0.7}\times10^9$ M$_\odot$, $\sigma_*$=300$\pm$50 km s$^{-1}$; Abell 3565 BCG, $M_{\rm bh}=1.3^{+0.3}_{-0.4}\times10^9$ M$_\odot$, $\sigma_*$=200$\pm$50 km s$^{-1}$. Figure 3 depict the $M_{\rm bh}$-$\sigma_*$ relations for the core elliptical galaxies and the other normal elliptical galaxies. The best-fit results are presented in Table 4. Two relations are still tight ($\epsilon_0\lesssim0.3$), consistent within 1$\sigma$ uncertainties, but the former is steeper ($\beta\simeq$ 5-6) than that later ($\beta\simeq4$). 
The mean values of the fitting results by $\chi^2$ estimator are
\beq
\alpha=8.23\pm0.11,\ \ \ \beta=5.09\pm0.80, 
\eeq
while the AB estimator gives a steeper slope ($\beta\simeq$6).

Our result confirms the prediction of Boylan-Kolchin et al. (2006) that the $M_{\rm bh}$-$\sigma_*$ relation for the core elliptical galaxies is steeper than that for the normal galaxies. 
The $M_{\rm bh}$-$\sigma_*$ relation derived from all the elliptical galaxies underestimates the mass of SMBHs in the center of core galaxies statistically. For the most massive galaxies ($\sigma_*\simeq400$ km s$^{-1}$), the discrepancy can be as large as factor 2$\sim$3.
The sub-sample of core galaxies overlap the one of the normal elliptical galaxies in the mass range of $10^8\sim10^{10}$ M$_\odot$, therefore the intrinsic scatter of the correlation for the combined sample in the high mass range is larger than that in the low mass range. 

\section{Conclusions and Discussion}

We examine whether early-type bulges and pseudobulges obey the same relation between black hole mass and stellar velocity dispersion, using the largest published sample to date. 
The main conclusions of this paper are as follows:

(1) On the relation for early-type bulges, we find a similar slope ($\beta=4.06\pm0.28$) and higher normalization ($\alpha=8.28\pm0.05$) than the previous result of Tremaine et al. (2002). The intrinsic scatter of the relation is $\lesssim$0.27 dex. 

(2) The pseudobulges have their own tight $M_{\rm bh}$-$\sigma_*$ relation with $\beta=4.5\pm1.3$, $\alpha=7.50\pm0.18$, which are different from the relation for early-type bulges over the 3$\sigma$ significance level. 

(3) The ``core" elliptical galaxies at high black hole masses end follow a steeper $M_{\rm bh}$-$\sigma_*$ relation with slope $\beta\simeq$ 5-6. 

The mass density of local SMBHs calculated with the  $M_{\rm bh}$-$\sigma_*$ relation (e.g. Yu \& Tremaine 2002; Wyithe \& Loeb 2003; Shankar et al. 2004) is consistent with the cumulative mass density of optical or X-ray selected quasar remnants expected from energy arguments (e.g. Soltan 1982; Fabian \& Iwasawa 1999; Yu \& Tremaine 2002; Shankar et al. 2004; Marconi et al. 2004). 
However, the $0.15\pm0.08$ dex higher normalization of our best-fit relation for the early-type bulges than the value of T02 will enhance the predicted mass and mass density of local SMBHs by $\sim40\%$.
In addition, the steeper slope of the relation for the core galaxies will raise the mass function of the most massive black holes in the universe, while the lower relation for the pseudobulges will reduce the mass function of the low mass SMBHs in the centers of disk galaxies.
Such modification will also affect estimation of the fraction of obscured AGNs, which can be tested independently by the observations of cosmic hard X-ray background. We will elaborate these implications in the subsequent papers.

We note unbiased estimate for parameters $\beta$ and $\alpha$ requires larger unbiased sample (both in $M_{\rm bh}$ and $\sigma_*$), or one must determine the selection bias carefully (e.g. Lauer et al. 2007b). 
However, the significant lower normalization ($\simeq$ 0.8 dex) of the $M_{\rm bh}$-$\sigma_*$ relation for pseudobulges seems not due to the effect of selection bias. At a given velocity dispersion, the SMBHs with massive $M_{\rm bh}$ are easier to be discovered and measured, because their influence spheres is larger. 
Our findings can be tested in larger unbiased samples, especially for the pseudobulges. The preliminary results of the $M_{\rm bh}$ upper limit on three SMBHs in pseudobulges\footnote{They are identified as pseudobulges by the nuclear rings for NGC 3351 and NGC 4314 (Shaw et al. 1995), and by the nuclear spirals for NGC 4143 (Martini et al. 2003).} (NGC 3351, NGC 4143, NGC 4314) indicate they also lie below the $M_{\rm bh}$-$\sigma_*$ relation for early-type bulges (Corsini et al. 2007).

Another interesting and unexpected finding is the agreement between the slope of $M_{\rm bh}$-$\sigma_{\rm e}$ relation and of the $M_{\rm bh}$-$\sigma_{\rm c}$ relation. T02 concluded that the earlier found discrepancy arises mostly from the different definitions of stellar velocity dispersion, but this difference in our sample is insignificant (cf. Figure 1). Our estimate of $\sigma_{\rm c}$ is much shallower than the previous results of FM00 and FF05. 
This ``converged'' slope will be important, and in some senses, convenient for the theoretical models and numerical simulations that try to reproduce the relation.

\begin{figure}
\centerline{\includegraphics[width=8.5cm]{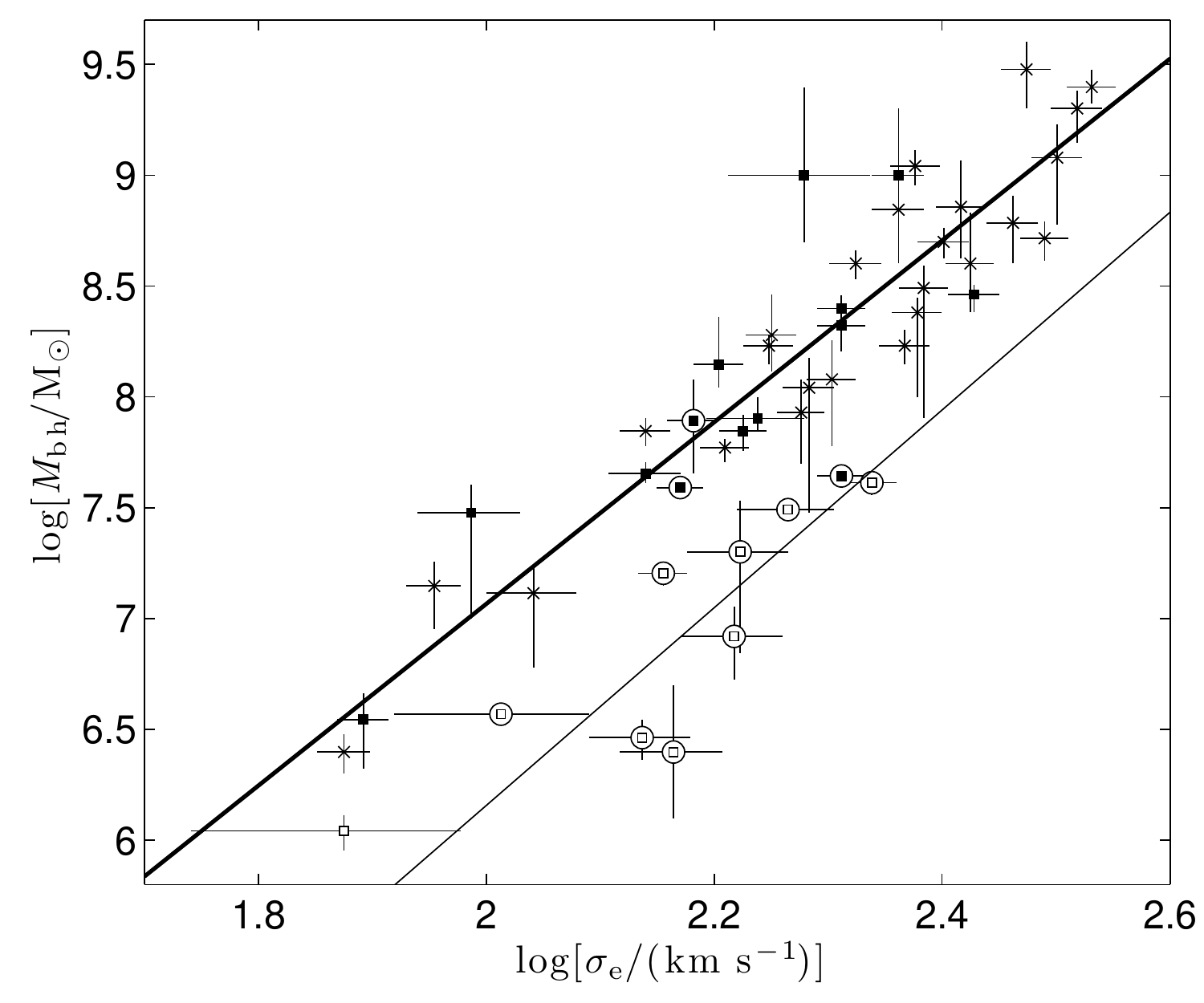}} 
\caption{Comparison of the $M_{\rm bh}$-$\sigma_{\rm e}$ relation for elliptical galaxies (stars), classical bulges (filled squares), and pseudobulges (open squares). The barred disk galaxies are marked by circles. The thick and thin solid lines are the best fit results for the early-type bulges and the pseudobulges respectively.}
\end{figure}

We compared the $M_{\rm bh}$-$\sigma_*$ relation for elliptical galaxies, classical bulges, pseudobulges, and barred/barless disk galaxies. As shown in Figure 4, elliptical galaxies and classical bulges follow the same $M_{\rm bh}$-$\sigma_*$ relation. In our sample, barred/barless distinction is only a result of classical/pseudobulges dichotomy. We also compared the inclination of the disk galaxies with classical bulges and pseudobulges (Figure 5). No obviously different distribution is found in our sample, the inclination seems not be responsible for the different $M_{\rm bh}$-$\sigma_*$ relations. 
However, we need larger sample to consider the effect of bar and inclination on the $M_{\rm bh}$-$\sigma_*$ relation.

\begin{figure}
\centerline{\includegraphics[width=8.5cm]{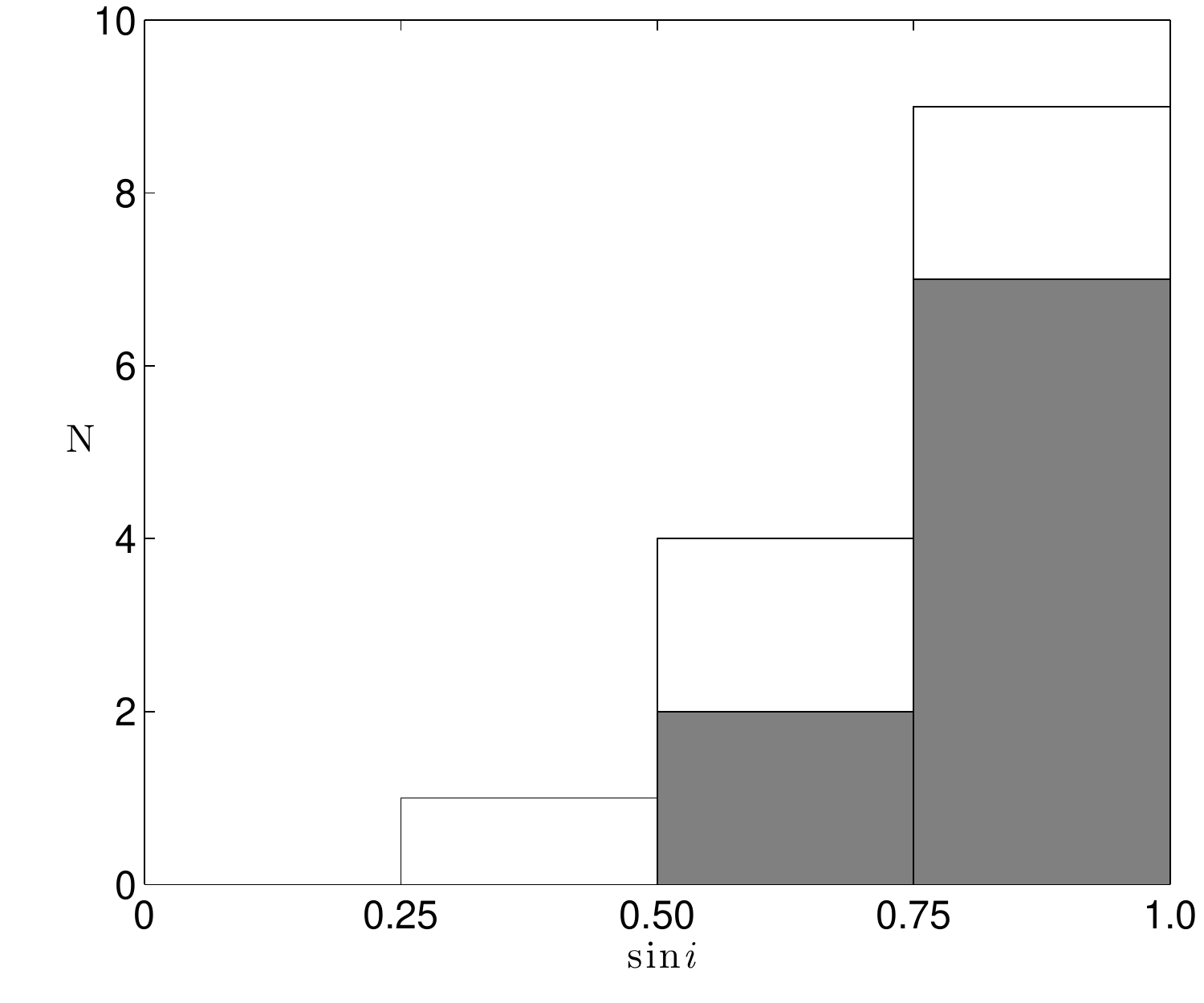}} 
\caption{Comparison of the inclination distribution of the disk galaxies with classical bulges (open columns) and pseudobulges (grey columns).}
\end{figure}

In stead of the assumed universal  $M_{\rm bh}$-$\sigma_*$ relation, the real correlation is composed of three sub-samples of SMBHs: hosted by  pseudobulges, by normal elliptical galaxies or classical bulges, and by ``core" elliptical galaxies. They have different slope and normalization, but similar intrinsic dispersion. 
The tight correlation between $M_{\rm bh}$ and $\sigma_*$ demonstrate that the bulge formation and SMBHs fueling are closely connected, despite the type of bulges. The strong feedback due to wet major mergers is the favored mechanism responsible for the formation of the correlation for the early-type bulges, while the slight steeper correlation for the core galaxies is build by the dry mergers of their early type progenitors. If the SMBHs in pseudobulges grow in the same period of their host bulge formation, their tight correlation may be the result of the secular evolution. The similar slope of the relations for early-type bulges and pseudobulges indicates they may experience the similar AGN feedback processes (e.g. Silk \& Rees 1998; King 2003), but the SMBHs growth in pseudobulges is relatively slower than that in the early-type bulges. We suggest it is due to the lower efficiency of AGN fueling by secular evolution. 
The other possibility is that the formation of these SMBHs is earlier than their host pseudobulges. The newborn disk galaxies may have the same $M_{\rm bh}$-$\sigma_*$ relation with that of elliptical galaxies, but the secular evolution steadily enhanced the central velocity dispersion, and the subsequent growth of SMBHs is insignificant. Therefore, the pseudobulges move rightward in the $M_{\rm bh}$-$\sigma_*$ diagram and deviate from the original relation.
In conclusion, the SMBHs formation and evolution scenario in the pseudobulges deserve further exploration, especially for the early stage of disk galaxy formation.

\section*{Acknowledgments}

I wish to thank Rashid Sunyaev for stimulating this work.
I am especially grateful to John Kormendy for helping to identify the bulge type of some galaxies.  
I thank Niv Drory, Peter Erwin and an anonymous referee for very useful comments.
Extensive use was made of the HyperLeda database, and the NASA/IPAC 
Extragalactic Database (NED), which is operated by the Jet Propulsion 
Laboratory, California Institute of Technology, under contract with NASA. 

\bsp
\label{lastpage}
 
\end{document}